\documentclass{sig-alternate-05-2015}

\usepackage{booktabs}
\usepackage{amsmath}
\usepackage{xfrac}
\usepackage{fixltx2e}
\usepackage{textcomp}
\usepackage{siunitx}
\usepackage{graphicx}
\PassOptionsToPackage{<pdftex>}{graphicx}
\usepackage{multirow}
\usepackage{color, soul}
\usepackage{subcaption}
\usepackage{caption}

\usepackage{enumitem}
\newcommand{\MTJ}[1]{MTJ\textsubscript{#1}}

\makeatletter
\def\@copyrightspace{\relax}
\makeatother

\begin{document}

\title{A Study on Performance and Power Efficiency of Dense Non-Volatile Caches in Multi-Core Systems}

\numberofauthors{1} 
%
\author{
	%
	%
	\alignauthor
	Amin Jadidi, Mohammad Arjomand, Mahmut T. Kandemir, Chita R. Das\\
	\affaddr{School of Electrical Engineering and Computer Science, Pennsylvania State University, USA }\\
	\email{\{axj945,mxa51,kandemir,das\}@cse.psu.edu}
}

\maketitle

In this paper, we present a novel cache design based on \emph{Mult-Level Cell Spin-Transfer Torque RAM} (MLC STT-RAM) that can dynamically adapt the set capacity and associativity to use efficiently the full potential of MLC STT-RAM.
We exploit the asymmetric nature of the MLC storage scheme to build cache lines featuring heterogeneous performances, that is, half of the cache line are read-friendly, while the other is write-friendly.
Furthermore, we propose to opportunistically deactivate ways in underutilized sets to convert MLC to \emph{Single-Level Cell} (SLC) mode, which features overall better performance and lifetime.
Our ultimate goal is to build a cache architecture that combines the capacity advantages of MLC and performance/energy advantages of SLC.
Our experiments show an improvement of 43\% in total numbers of conflict misses, 27\% in memory access latency, 12\% in system performance (i.e., IPC), and 26\% in L3 access energy, with a slight degradation in cache lifetime (about 7\%) compared to an SLC cache.

\printccsdesc


\section{Introduction}
\label{Intro}
Ever increasing number of cores in chip multiprocessors (CMPs) coupled with the trend toward rising working set sizes of emerging workloads (such as web-search algorithms, social networking and modern databases) stresses the demand for large, multi-level, on-die cache hierarchy to hide long latency of off-chip main memory.
During the last few decades, SRAM-based cache memories successfully kept pace with this capacity demand by exponential reduction in cost per bit. However, entering sub-20nm technology era, where leakage power becomes dominant, it is very challenging to continue with the expansion of the cache hierarchy while avoiding the power wall. 
One promising approach to address this problem is to replace SRAM with a non-volatile memory (NVM) technology~\cite{Y1,Dong08:STT-RAM-Circuit}. 
Recently, various NVMs have been prototyped in nano-technology regime~\cite{Driskill:STT-RAM,Ishigaki:MLC-VLSIT} and many of them are expected to be commercially available by the end of the decade~\cite{Qureshi2010}. Among such memory technologies, Spin-Transfer Torque RAM (STT-RAM) is the best candidate for use within the processor; STT-RAM has zero leakage power, accommodates almost 4$\times$ more density than SRAM, and has small read access latency and high endurance.

The storage element in a STT-RAM cell is a Magnetic Tunnel Junction (MTJ) which stores binary data in form of either parallel magnetic direction (set) or anti-parallel magnetic direction (reset). 
Two types of STT-RAM cell prototypes can be realized: \emph{Single-Level Cell (SLC)} STT-RAM and \emph{Multi-Level Cell (MLC)} STT-RAM. 
The SLC STT-RAM cell consists of one MTJ component which is used to store one bit information.
The MLC STT-RAM device, on the other hand, is typically composed of multiple MTJs, which are connected either serially or in parallel, and are used to store more than one bit information in a single cell. 
Such increased density in MLCs comes at the cost of linear increase in access latency and energy with respect to the cell storage level (i.e., the number of bits stored). For instance, the read (or write) latency and energy consumption of a 2-bit STT-RAM cell is two times higher than that of a SLC STT-RAM device under same fabrication technology. Furthermore, MLC STT-RAM usually has lower endurance (in terms of write cycles) compared to the SLC.
In short, SLC and MLC storage elements show two different characteristics: SLC is fast, power-efficient, and has a long lifetime; but, MLC trade-offs these metrics for high density.

Over the past few years, several device-level and architect\-ure-level optimizations/solutions has been proposed that attempt to address the issues of high write latency/energy~\cite{Y1,Dong08:STT-RAM-Circuit,Mishra11:3D-STT-RAM,Chen12:Hybrid-STT-RAM,Sun09:STT-RAM,Sun11:STT-RAM-Retention} and limited endurance~\cite{i2wap,Sequoia} in SLC STT-RAM caches. 
However, little attention has been paid to explore the potential of the MLC STT-RAM cache in multicore systems.
Indeed, such an analysis and study is necessary as feature scaling is continuing and employing MLC devices seems to be the only way of increasing cache capacity in a cost-efficient and power-efficient manner.
In this paper, we focus on the MLC STT-RAM cache design space exploration and related optimizations, when it is used as last level cache (LLC) in CMPs. 
Specifically, this paper tries to answer the following research questions:
\begin{enumerate}
\item What is the design of an MLC STT-RAM cache? Can we reduce its read and write access latency/energy by employing logic-level optimizations?
\item What are the performance implications of an MLC STT-RAM cache compared to an SLC STT-RAM cache in an iso-area design? What kinds of workloads can get benefits from MLC STT-RAM cache configuration? And what kinds of workloads exhibit performance degradation in a system with MLC STT-RAM cache? 
\item What types of architecture-level optimizations can be employed to further improve the performance-efficiency and energy-efficiency of a MLC STT-RAM cache? After all logic-level and architecture-level optimizations, does the resultant MLC STT-RAM cache perform better than its SLC counterpart for a wide range of workload categories?
\end{enumerate}

This paper answers the above research questions in detail and introduces the following novel mechanisms to tackle the challenges brought by an MLC STT-RAM cache:
\begin{itemize}
\item To reduce the read/write access latency and energy of an MLC-based cache, we introduce \emph{stripped data-to-cell mapping} scheme as a logic-level optimization. This design mainly relies on the asymmetric behavior of reading or writing different bits of an MLC; while one bit can be read fast, using the other saves more energy during writes. Instead of storing cache lines next to one another in independent memory cells, this mapping enables data blocks to be ``stacked'' on top of each other -- storing bits of two cache lines in the same cell, one as MSB and the other as LSB. With this data layout arrangement, we demonstrate that, for half of the cache lines, the read/write access latency is comparable with SLC cache; and for the rest of the cache lines, the read/write access energy is in range of an SLC cache.
\item To further improve performance of the cache, we propose an \emph{associativity adjustment scheme}. This scheme tries to adjust the associativity degree of each set independently by switching on or off the ways stacked onto others. With this mechanism in place, each set cache acts like highly-associative cache for sets that benefit from more associativity and behaves like a low-associative cache when extra capacity is not useful for a set, thereby reducing both energy consumption and access latency. The main feature of this dynamic design is that it can easily determine cache associativity with limited hardware overhead.
Indeed, this scheme only requires (1) a mechanism to detect workload behavior within each set and (2) a read and write-aware inter-set data movement which does not alter any address and data path in the cache hierarchy.
\item We also propose a swapping policy to enhance performance and energy of the cache, on top of the other two schemes. When used with a stripped cache design, this scheme tries to place the read-dominant cache blocks to the ways with small read access latency, while it places the write-dominant cache blocks to the ways with small write energy.
\end{itemize}

Compared to a \emph{Single-Level Cell} (SLC) STT-RAM cache with limited associativity, the proposed design reduces conflict misses of the sets with large large working set.
And, compared to a MLC STT-RAM cache with conventional stacked data-to-cell mapping, it improves read performance and write energy by converting MLC lines to SLC when a set does not need extra cache capacity.

\begin{figure}[!t]
  \center
  \includegraphics[width=8cm, height=6.3cm]{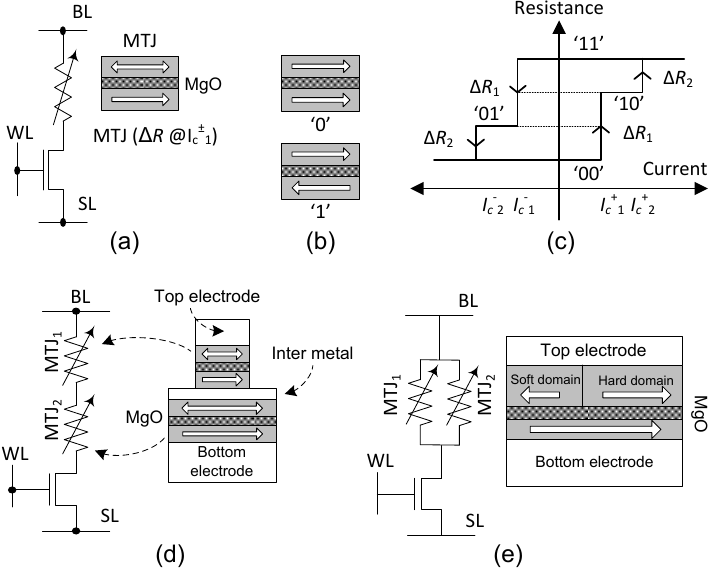}
  \caption{(a) SLC STT-RAM cell consisting of one access transistor and one MTJ storage carrier (``1T1J''); (b) Binary states of an MTJ: two ferromagnetics with anti-parallel (or parallel) direction indicate a logical `1' (or `0') state; (c) MLC STT-RAM cell with serial MTJs: the \emph{soft-domain} on top of the \emph{hard-domain}; (d) Resistance levels for 2-bit STT-RAM: four resistance levels are obtained by combining the states of two MTJs having different threshold current.}
  \label{fig-STT}
\end{figure}

\section{Overview of STT-RAM}
\label{Back2}

\subsection{Single-Level Cell (SLC) Device}
STT-RAM is a scalable generation of Magnetic Random Access Memory (MRAM).
Figure~\ref{fig-STT}a shows the basic structure of an SLC STT-RAM cell which is composed of a standard NMOS access transistor and a \emph{Magnetic Tunnel Junction} (MTJ) as the information carrier -- This forms a ``1T1J'' memory element.
Each MTJ consists of two ferromagnetic layers and one oxide barrier layer.
One of the ferromagnetic layers (i.e., the \emph{reference}) has a fixed magnetic direction and the magnetic direction of the other (i.e., the \emph{free} layer) can be changed by passing a spin-polarized current into the cell.
If these two ferromagnetic layers have anti-parallel (or parallel) direction, the MTJ resistance is high (or low), indicating a binary value of `1' (or `0') state (Figure~\ref{fig-STT}b).
Compared to conventional MRAM, STT-RAM exhibits superior scalability since the threshold current required to make the status reversal (from `0' to `1' or vice versa) decreases as the MTJ size shrinks.
Furthermore, the STT-RAM technology has reached a good level of maturity and the fabrication cost of cells in this technology is small -- the number of additional mask steps beyond a standard CMOS process is no more than two (less than 3\% added cost)~\cite{Driskill:STT-RAM}.

\subsection{Multi-Level Cell (MLC) Device}
The MLC capability can be realized by modeling four or more resistance levels within one cell. 
With the current VLSI technologies, the MLC STT-RAM devices are fabricated by using multiple MTJs structured either in parallel or series connections.
Figure~\ref{fig-STT}c depicts these two configurations for a 2-bit MLC device -- the 2-bit series STT-RAM is composed of two vertically stacked MTJs, and the 2-bit parallel cell is composed of one common reference layer and two independent free layers which are arranged side-by-side (parallel).
Under the same fabrication conditions and process variation, the serial MLC has advantages over the parallel design: (1) it exhibits much lower bit error rates related to read and write disturbance issues~\cite{Zhang:MLC-ICCAD}; (2) it results in smaller cells; and (3) while it has full-compatibility with modern perpendicular STT-RAM technology, fabricating parallel MTJs with this technology is very challenging.
We consider serial MTJ-based 2-bit STT-RAM cells in this work, although our ideas apply equally to parallel MTJ-based devices and MLCs with higher bit densities.

In the 2-bit serial STT-RAM, MTJs have different layer thickness and area. This results in different threshold current, $I_c^{\pm}$ and resistance variation, $\Delta{R}$ for the two MTJs.
Four levels of resistance are obtained by combining the binary states of both MTJs (Figure~\ref{fig-STT}d).
In such a cell, the layer that requires a small current to switch (MTJ\textsubscript{1}) is referred to as \emph{soft-domain} and the layer that requires a large current to switch (MTJ\textsubscript{2}) is referred to as \emph{hard-domain}.
Assuming 2-bit information, the \emph{Least Significant Bit} (LSB) and the \emph{Most Significant Bit} (MSB) are stored into the \emph{soft-domain} and \emph{hard-domain}, respectively.
Below, we describe details of the write and read operations for a 2-bit series MLC.

\subsubsection{Two-step Write Operation} 
The circuit schematic for access operations on a 2-bit serial STT-RAM is depicted in Figure~\ref{fig-MLC-STT}.
Two MTJs are accessed by an access transistor controlled by a word-line (WL) signal and a write current always passes through both MTJs.
Accordingly, when the \emph{hard-domain} is written, the state of the \emph{soft-domain} is also switched into the same direction, because of its larger threshold current ($I_c$).
In order to write the LSB, the direction and amplitude of the current pulse are defined by the logical value of the MSB and LSB as well as the characteristics of both MTJs.
If the desired LSB value is equal to the currently stored MSB, a second current pulse is not necessary, since the \emph{soft-domain} was already switched to the proper state.
Otherwise, a second current pulse is required to write the LSB into the \emph{soft-domain} (Figure~\ref{fig-MLC-STT}c).
This second current pulse should be set between the $I_c$ values of the two MTJs to prevent bit flip of the \emph{hard-domain} (Figure~\ref{fig-MLC-STT}b).
Consequently, the number, direction, and amplitude of write current pulses in this two-step write scheme vary depending on the written data.

\subsubsection{Two-step Read Operation} 
\begin{figure}[!t]
  \center
  \includegraphics[width=3.2in]{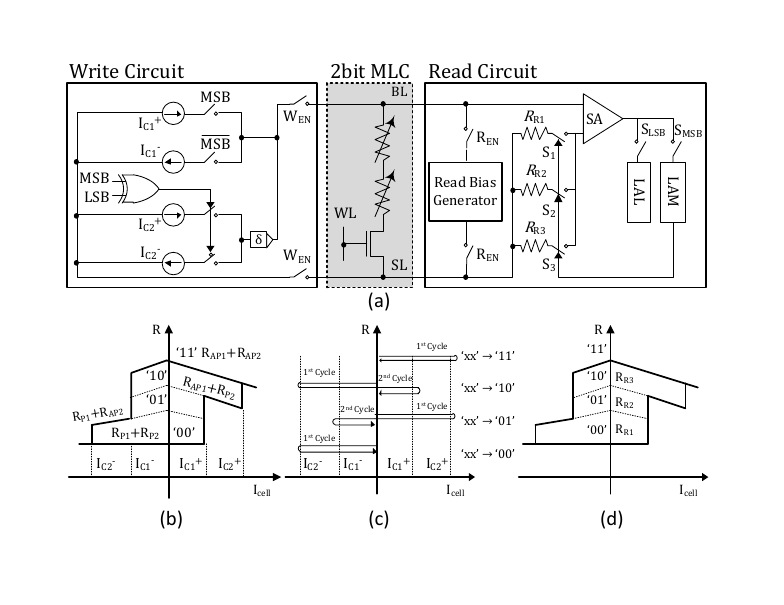}\vspace{-10pt}
  \caption{(a) Schematic of the read and write access circuits in a 2-bit MLC cache array; (b,c) write operation transition model: first, the MSB is written to the \emph{hard-domain} and if the LSB differs from MSB a small current is driven to switch its direction; (d) three resistance references in a sense amplifier, each between the two neighboring resistance states.}
  \label{fig-MLC-STT}
\end{figure}

On a read, the access transistor is turned on and a voltage difference is applied between the bit-line (BL) and source-line (SL) terminals.
This voltage difference causes a current to pass through the cell and is small enough to avoid any MTJs to switch its magnetic direction.
The value of the current is a function of the MTJs' resistance and is input into a multi-reference sense amplifier.
The sense amplifier unit has three resistance references, each between two neighboring resistance states (Figure~\ref{fig-MLC-STT}d).
The cell content is read through two read cycles in a binary search fashion.
In the first step, the sense amplifier uses R\textsubscript{2} as a reference to identify the value of MSB stored in the \emph{hard-domain}.
In the second step, depending on the MSB value, the reference resistance is switched to R\textsubscript{1} or R\textsubscript{3}, and the LSB is read from the \emph{soft-domain}.

\begin{table}[!t]
  \centering
  \caption{2-bit MLC STT-RAM compared to SLC cell circuit model~\cite{Ishigaki:MLC-VLSIT} at the 45~nm PTM.}
  \label{table-model}
    \scriptsize
    \begin{tabular}{|p{0.73in}||c|c||c|}
    \hline
    \multirow{2}{*}{\textbf{Parameters}}  & \multicolumn{2}{c||}{\textbf{MLC}}    & \multirow{2}{*}{\textbf{SLC}}\\
    \cline{2-3}
                    & LSB (\MTJ{1})    & MSB (\MTJ{2})    & \\
      \hline
      Dimensions    & 70$\times$140~nm & 75$\times$150~nm & 75$\times$140~nm \\
      \hline \hline
      Read Latency  & 0.962~ns         & 0.962~ns         & 0.856~ns \\
      Read Energy   & 0.0115~pJ        & 0.0115~pJ        & 0.0112~pJ \\
      \hline \hline
      Write Latency & 10~ns            & 10~ns            & 10~ns  \\
      Write Energy  & 1.92~pJ          & 3.192~pJ         & 3.192~pJ \\
      \hline \hline
      Endurance & \multicolumn{2}{c||}{10\textsuperscript{12} Writes} & 10\textsuperscript{10} Writes \\
      \hline
    \end{tabular}
\end{table}

\vspace{-15pt}
\subsection{SLC versus MLC: Device-Level Comparison}
\label{new-model}
Table~\ref{table-model} gives the typical latency, energy and lifetime parameters of a 2-bit MLC and a SLC STT-RAM cell.
These parameters are taken from state-of-the-art prototypes (e.g.,~\cite{Ishigaki:MLC-VLSIT}) at the 45nm PTM technology. Based the parameters in this table: we can compare SLC and MLC in four ways:
\begin{itemize}
\item \textbf{Cell Area --} In STT-RAM, the MTJ size is larger than the access transistor and determines the cell size. As a result, under same technology constraints, the SLC has the same area as smaller MTJ of the MLC (MTJ\textsubscript{1}), i.e., 70$\times$140, and MLC area is 75$\times$150~nm which is a little larger (due to larger size MTJ\textsubscript{2}). As such, in an ISO-area design, the capacity of the 2-bit MLC cache is less than two times of the SLC cache.
\item \textbf{Read Operation --} Both devices use the same read voltage difference (i.e., -0.1~V~\cite{Ishigaki:MLC-VLSIT}), which results in a read latency of 0.856ns for SLC cell, and total read latency of 1.9ns for the MLC cell (about 0.9~ns per bit which is in the same range of SLC). The same discussion can be made for energy consumption.
\item \textbf{Write Operation --} In order to meet the write performance requirement of LLC, the writing pulse width is set to 10~ns~\cite{Chen10:MLC-STT} (which is in range of today's SRAM caches).
For 2-bit MLC, this results in the write energy of the MTJ\textsubscript{2}(\emph{soft-domain}) and MTJ\textsubscript{1}(\emph{hard-domain}) to be 1.92pJ and 3.192pJ, respectively~\cite{Lou08:AppliedPhy}. Also, writing into SLC cell consumes the same energy as writing into MTJ\textsubscript{2}, as both employ the same sized MTJs.
\item \textbf{Cell Endurance --} Due to the need for larger write currents and two-step write operation in the MLC cell, it generally has lower cell endurance than SLC (about 100$\times$ based on the model in Table~\ref{table-model}). 
\end{itemize}

Throughout this paper, we use the values in this table for performance, energy and endurance analysis of the STT-RAM-based cache designs.

\section{MLC STT-RAM Cache: The Baseline}
\begin{figure}[!t]
  \center
  \includegraphics[width=\columnwidth]{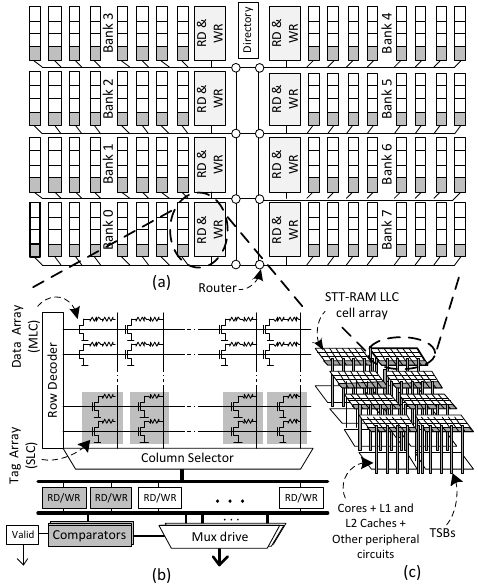}
  \caption{2-bit STT-RAM layout and a schematic view of the cache array, read and write circuits.}
  \label{fig-STT-Array}
\end{figure}

We assume a chip multiprocessor (CMP) with an MLC-based STT-RAM last-level cache (LLC). 
Figure~\ref{fig-STT-Array} shows the architectural model of this system where, due to CMOS non-compatibility  of STT-RAM, the LLC is integrated into the processor die using 3D VLSI\footnote{Today, 3D ICs are commercially available~\cite{hmc} and have been receiving immense research interest from early 2000. Besides its latency and bandwidth benefits, one of the advantages of 3D ICs over 2D ICs is that they provide a platform to integrate different (non-compatible) technologies on the same die, with less concern on impacts of noises and fabrication costs.}. This system is a 2-tier 3D chip.
At the processor tier, cores and all lower caches are placed. 
The STT-RAM cache (LLC), which is at the upper layer, is logically shared among all cores while physically structured as static NUCA and mounted at the top tier of 3D die. Note that prior studies assume the same integration model for NVM caches in future processors~\cite{Sun09:STT-RAM,Black06:3D,Mishra11:3D-STT-RAM}.
The modeled STT-RAM cache has two characteristics, namely, \emph{NUCA structure} and \emph{serial-loop access}, which are explained below.

\textbf{NUCA structure --} The LLC in our CMP model is a large cache, which has to be structured as a NUCA (Non-Uniform Cache Architecture) for scalability and high-perfor\-mance. 
In NUCA, a large cache is divided into multiple banks (one bank per each core) connected by an on-chip network for data and address transfer between the banks.
NUCA exhibits non-uniform access latencies depending on the physical location of the cache bank being accessed~\cite{Muralimanohar07:NUCA-Interconnect}.
Two main types of NUCA have been proposed: static NUCA and dynamic NUCA.
In static NUCA, a cache line is statically mapped into banks, with the lower bits of the index determining the bank.
In dynamic NUCA, on the other hand, any given line can be mapped into several banks based on a placement policy.
Although dynamic NUCA can drastically reduce average access latency (compared to the static NUCA) by putting the frequently-accessed data into banks that are closer to the source of request~\cite{Kim02:NUCA}, it is not a good design choice in non-volatile caches (like our design), since it increases the cache write traffic (due to frequent movement of cache lines between banks), which can in turn accelerate the wear-out problem.
Therefore, we assume that the LLC is built as a static NUCA.

\textbf{Serial-lookup access --}
The modeled LLC is a serial-lookup cache.
In serial caches, tag and data arrays are accessed sequentially, saving energy at the expense of increased delay.
The serial cache access latency relies heavily on the tag array latency, and consequently, we choose SLC for the tag array to minimize the latency.
Figure~\ref{fig-STT-Array}b shows the modeled LLC with SLC STT-RAM tag array and 2-bit MLC STT-RAM data array.
As shown in this figure, STT-RAM has the same peripheral interfaces used in SRAM caches: each bank consists of a number of cache lines, decoders for set index, read circuits (RDs), write circuits (WRs), and output multiplexers.
Unlike SRAM, however, the current sense amplifiers in the STT-RAM read circuit are shared and multiplexed across bit-lines due to their large size compared to the cell array.
For the read and write operations, a decoder circuit selects a cache set and connects the selected physical line to RDs for reading or WRs for writing.

\subsection{Stripped Data-to-Cell Mapping}
\label{Stripping}
\begin{figure}[!t]
	\center
	\includegraphics[width=\columnwidth]{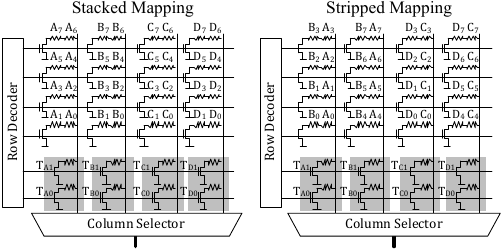}
	\caption{An illustration of stacked versus stripped data-to-cell mapping for a 8-bit data array (four 2-bit MLC cells) and 2-bit tag arrays (in SLC).
		In stacked data-to-cell mapping scheme, data bits of the same cache block (each one has 8 bits) are mapped to 4 independent memory cells (i.e., 2 in each cell). 
		In stripped mapping, each memory cell contains only one bit of each cache block -- so, each 8-bit cache block spans over 8 memory cells. For instance, lines `B' and `D' are mapped to \emph{hard-domains} (MTJ\textsubscript{1}), whereas `A' and `C' use \emph{soft-domains} (MTJ\textsubscript{2}).}
	\label{fig-mapping}
\end{figure}

In the discussed 2-bit MLC model (Section~\ref{new-model}), both bits are assumed to be written or read together, although sequentially.
Applied to a cache context, both bits of a cell would normally be mapped to the same cache line, as illustrated on the left side of Figure~\ref{fig-mapping}.
In this \emph{stacked} data-to-cell mapping, reading a cache line always takes two read cycles (i.e., 1.9ns in our 2-bit model), while writing takes two write cycles at most (i.e., 20ns). In other words, with the stacked mapping, the access latency to an MLC STT-RAM cache is roughly twice that of an SLC cache. The same discussion applies to the energy consumption of each access.
As opposed to this design, we propose to exploit the read and write asymmetry of the two MTJs in order to simultaneously optimize overall access latency and energy consumption. 

As discussed earlier, when reading or writing bits separately in a 2-bit STT-RAM, we observe a different latency and energy consumption: for read operations, the MSB can be read from the \emph{hard-domain} in a single read cycle (0.96~ns for the 2-bit model in Table~\ref{table-model}), whereas reading the LSB requires a second read cycle.
For write operations, on the other hand, one can write into either the \emph{hard-domain} or the \emph{soft-domain} independently, by using a single current pulse.
Unlike the \emph{soft-domain}, writing into the \emph{hard-domain} has two effects:
\begin{enumerate}
	\item Writing into the \emph{hard-domain} may cause the LSB to flip. Thus, each write request to MSB must be preceded with an LSB read, which takes an extra 1.9~ns (or generally two read cycles). Following that, both MTJs are written sequentially that is, first the MSB and then the read LSB.
	\item Writing into the \emph{hard-domain} dissipates 1.6 times the energy required for the \emph{soft-domain}. The larger is the writing current, the shorter the cell lifetime. 
\end{enumerate}
Accordingly, although the \emph{hard-domain} emulates SLC read access in latency, the cost of writing into the \emph{soft-domain} is much lower, primarily because only a small current is required to switch its polarity.

\begin{figure*}[!t]
    \centering
    \begin{subfigure}[t]{0.47\textwidth}
        \centering
        \includegraphics[width=\textwidth]{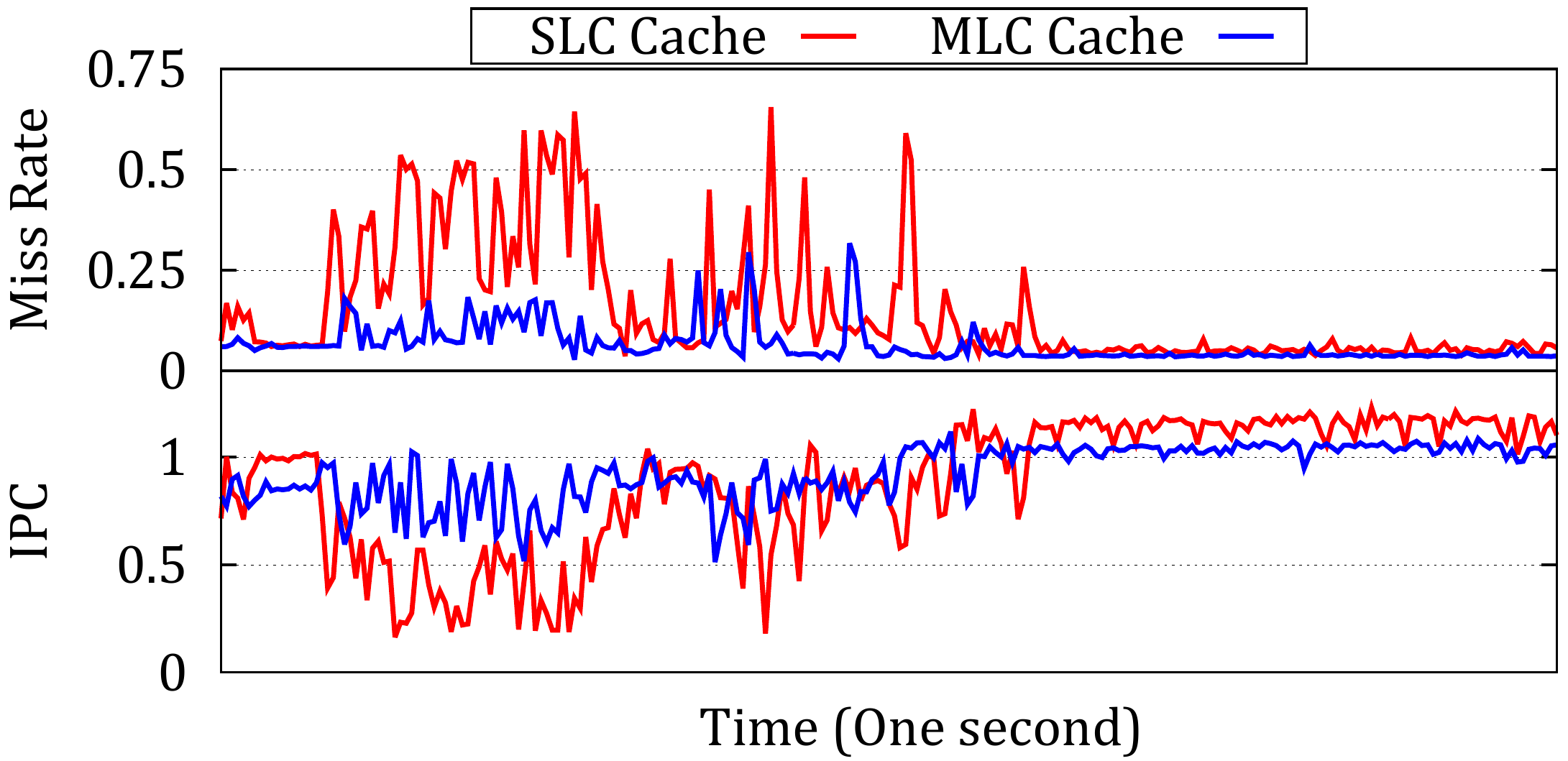}
        \caption{\emph{gcc}}
    \end{subfigure}
    ~ ~ ~
    \begin{subfigure}[t]{0.47\textwidth}
        \centering
        \includegraphics[width=\textwidth]{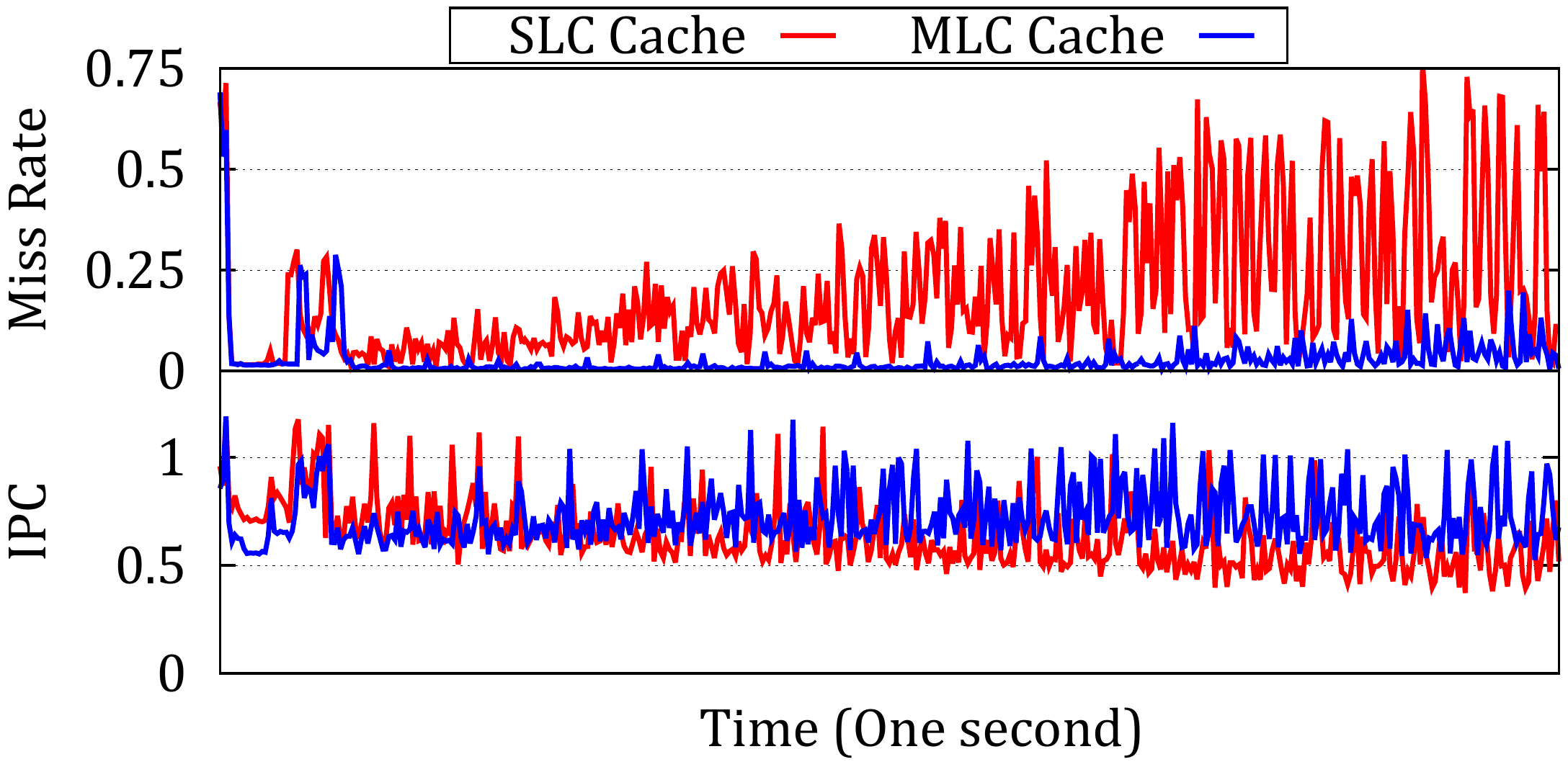}
        \caption{\emph{xalancbmk}}
    \end{subfigure}
    ~ ~ ~
    \begin{subfigure}[t]{0.47\textwidth}
        \centering
        \includegraphics[width=\textwidth]{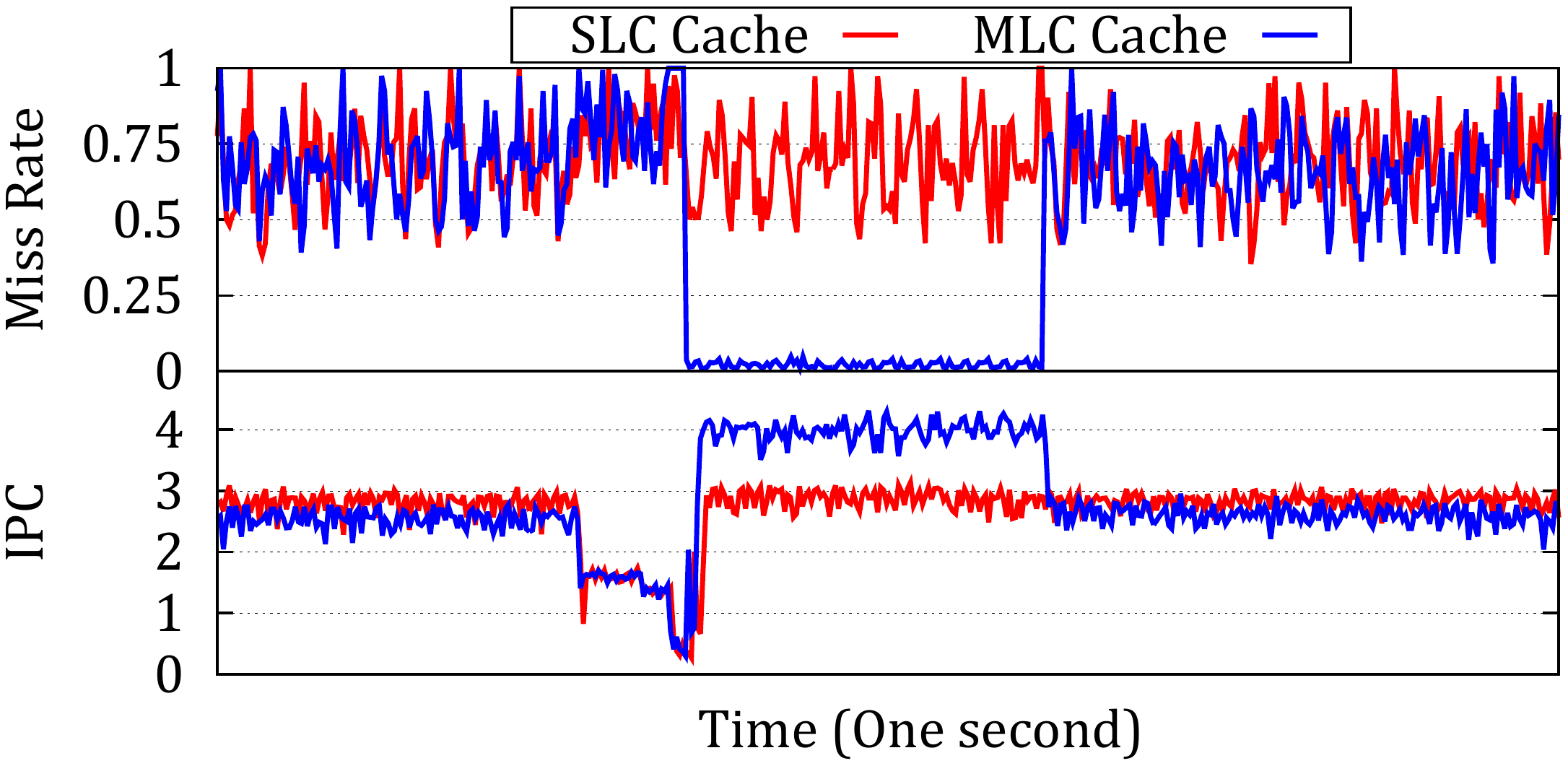}
        \caption{\emph{hmmer}}
    \end{subfigure}
    ~ ~ ~
    \begin{subfigure}[t]{0.47\textwidth}
        \centering
        \includegraphics[width=\textwidth]{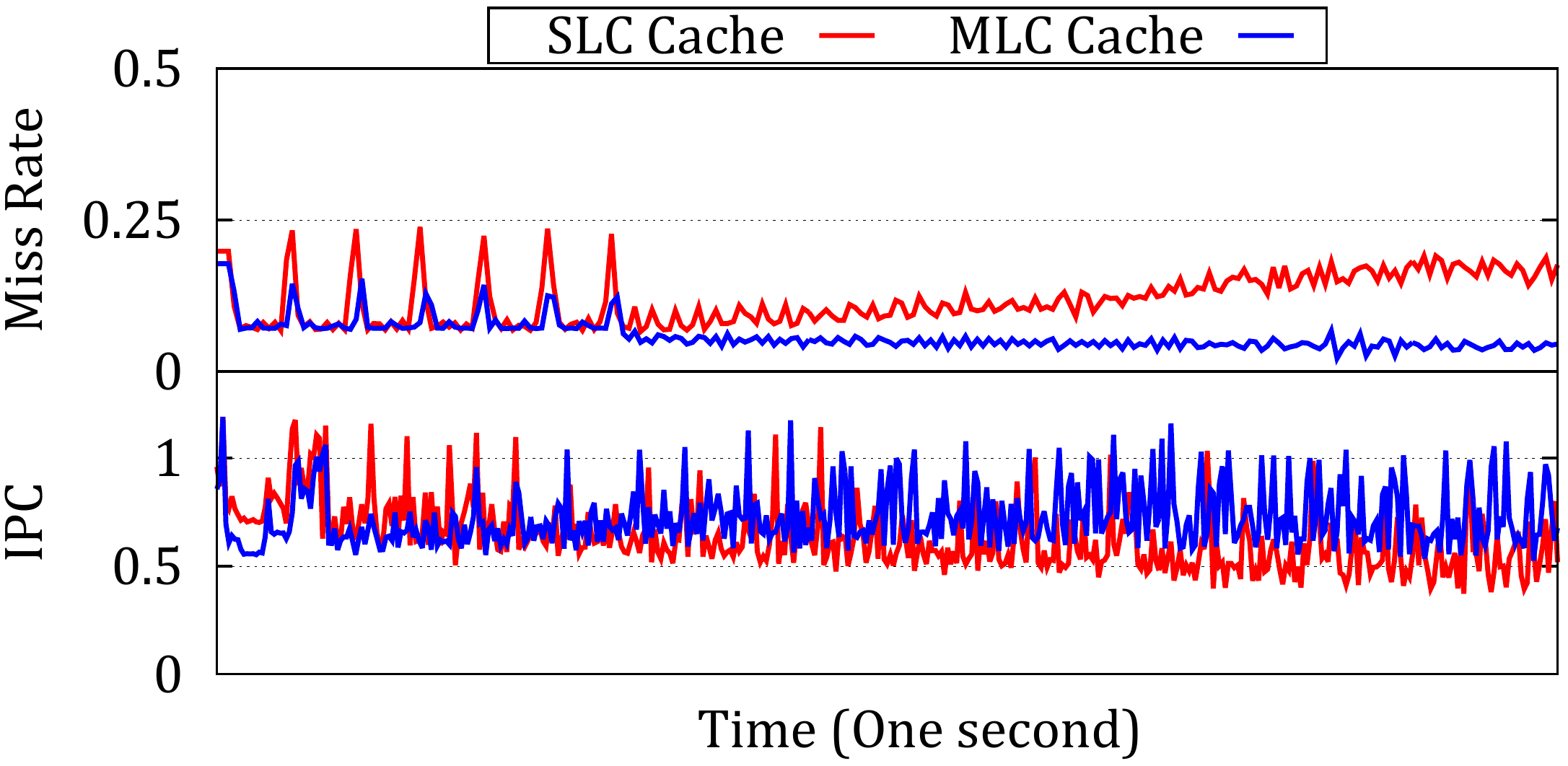}
        \caption{\emph{omnetpp}}
    \end{subfigure}    
    \caption{Performance comparison of the SLC and MLC-based STT-RAM caches in terms of the LLC miss ratio and IPC (as a system-level metric) for four workloads from the SPEC-CPU 2006 benchmark suite.}
    \label{fig:eval1}
\end{figure*}

\begin{figure}[!t]
  \center
  \includegraphics[width=\columnwidth]{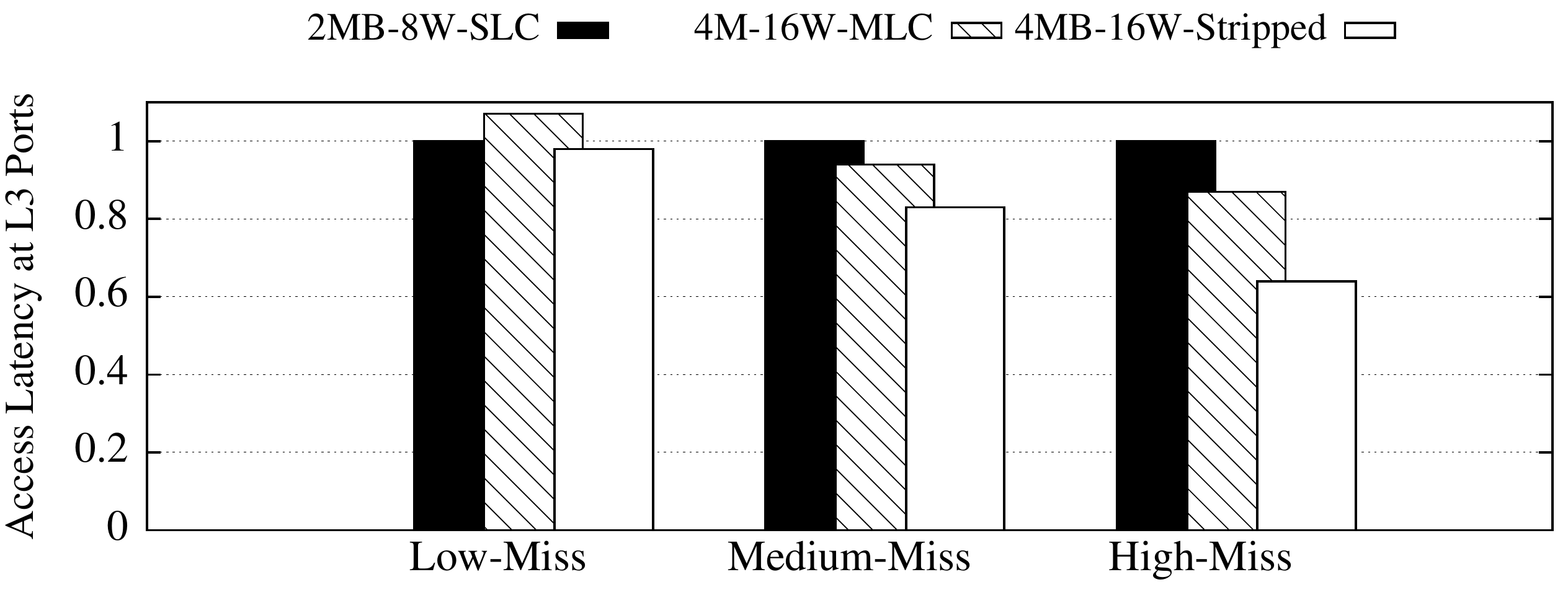}
  \caption{Comparison of the stripped cache with SLC array and the stacked 2-bit MLC, each with the same die area.
Stripped MLC is better than SLC in applications with \emph{high} and \emph{medium} L3 miss rates, since it increases the effective cache capacity in terms of lines and associativity on-demand.
It is also better than the conventional stacked MLC cache in applications with \emph{low} L3 miss rates as it constructs the \emph{fast read} lines.}
  \label{fig-ideal}
\end{figure}

Based on the different read/write characteristics of both MTJs, we propose a \emph{stripped} data-to-cell mapping, which groups the \emph{hard-domains} together to form \emph{Fast Read High-Energy write} (FRHE) lines, and groups the \emph{soft-domains} to form \emph{Slow Read Low-Energy write} (SRLE) lines.
The right portion of Figure~\ref{fig-mapping} depicts the logical arrangement of the tag and data arrays for the stripped mapping.
Within each cache set, half of the cache lines will be mapped to the \emph{hard-domains} and the other half to the \emph{soft-domains}.
Table~\ref{table-transactions} summarizes the sequence of transactions required to read and write the FRHE or SRLE lines in a stripped 2-bit cache.

\begin{table}[!b]
  \centering
  \caption{Sequence of transactions when accessing a 2-bit MLC cache with stripped data-to-cell mapping. FRHE (Fast Read High-Energy write) allocates \emph{hard-domain}s (\MTJ{2}), and SRLE (Slow Read Low-Energy write) allocates \emph{soft-domain}s (\MTJ{1}).}
  \label{table-transactions}
  \scriptsize
    \begin{tabular}{|p{0.7in}||p{2.25in}|}
      \hline
      \textbf{FRHE Read} & (1) Read \emph{hard-domain}s\\ \hline
      \textbf{FRHE Write} & (1) Read \emph{hard-domain}s; (2) Read \emph{soft-domain}s; (3) Write \emph{hard-domain}s; (4) Write \emph{soft-domains} \\ 
      \hline
      \textbf{SRLE Read} & (1) Read \emph{hard-domain}s; (2) Read \emph{soft-domain}s\\ 
      \hline
      \textbf{SRLE Write} & (1) Write \emph{soft-domain}s\\
      \hline
    \end{tabular}
\end{table}

In addition to its performance efficiency, the stripped data-to-cell mapping provides some other opportunities for optimizing energy consumption and lifetime of a 2-bit MLC cache. Before describing these advantages, we first evaluate the performance efficiency of a MLC-based cache (with stripped mapping) when running real workloads.


\subsection{Performance Analysis}
Employing an MLC STT-RAM cache has two opposing impacts on the performance of a multi-core system.
On the one hand, thanks to its large capacity, it can improve performance by reducing the cache miss rate and so the need for accessing off-chip main memory if employed as an LLC (like our model). This is especially true for emerging workloads (like social networking applications and new database workloads) that usually have very large working set sizes.
On the other hand, due to its high read and write access latencies, such a cache architecture degrades the system performance for workloads with low or negligible miss rates for on-chip caches.
Although the stripped data-to-cell mapping partially addresses the problem of high access latency, the problem still exists for half of the cache lines which always exhibit the maximum MLC STT-RAM read and write latencies.

To examine the performance efficiency of a 2-bit STT-RAM cache with the stripped data-to-cell mapping, Figure~\ref{fig:eval1} presents the \emph{LLC miss rate} and \emph{Instruction-per-Cycle (IPC)} of a single-core system for one-second execution of four typical applications from SPEC-CPU 2006~\cite{Spradling07:SPECCPU06}. 
The applications are chosen to cover a wide range of scenarios with the low, moderate, and high LLC miss rates. In this experiment, we skip the 100~Million instructions as the initialization stage, and the results for the next 200~million instructions are collected.
We assume two LLC configurations: (1) a 256KB SLC-based STT-RAM cache with 8 ways per each set, and (2) a 512KB 2-bit STT-RAM cache with 16 ways per each set and the proposed stripped data-to-cell mapping. In both the configurations, the cache line size is set to 64B. Note also that, both the configurations have almost the same LLC area size (i.e., this is an iso-area analysis)\footnote{For this experiment, we used the same simulation platform described in Section~\ref{Experiments}.}.

One can make two main observations from the results in this figure. \emph{First}, if an application exhibits low LLC miss rate during some of its phases or its entire execution, the SLC cache results in better performance (i.e., lower IPC), thanks to its lower access latency than the MLC-based cache. \emph{Second}, during the phases where SLC cache's miss rate is high, the MLC cache can increase IPC if it can hold the whole or major part of the working set size of the application. Indeed, there are the cases where the application's memory footprint is very large (even larger than MLC cache size) or the application has a streaming behavior (i.e., accessing a large set of addresses in a sequential fashion without reuse), during which both the SLC and MLC caches have high miss rates and consequently low IPC values.  
Note that we have observed the same behavior for energy consumption of the SLC and MLC caches, but due to lack of space, we will discuss the energy consumption results in the evaluation section.

To generalize our observation to large LLC sizes and evaluate the performance efficiency of the stripped data-to-cell mapping, Figure~\ref{fig-ideal} compares memory access latency of a 4MB 16-way stripped LLC (L3) with two extreme baselines: (1) a 2MB 8-way SLC cache with nearly the same die area (i.e., \emph{fast cache}), and (2) a 4MB 16-way 2-bit cache with stacked data-to-cell mapping (i.e., \emph{dense cache}). 
For this experiment, we use the same evaluation methodology described later in Section~\ref{Experiments} and the workload set in Table~\ref{table-Workload}.
This figure confirms that stripped MLC let us \emph{have cake and eat it too}: in applications with \emph{high} and \emph{medium} LLC miss ratios (first two columns), the performance improvement over the SLC baseline is mainly due to the increase in effective cache capacity; while in applications with \emph{low} L3 miss ratio, it reduces the access time of the MLC cache by constructing separate \emph{fast read} and \emph{write} lines.

\begin{figure*}[!t]
    \centering
    \begin{subfigure}[t]{0.47\textwidth}
        \centering
        \includegraphics[width=\textwidth]{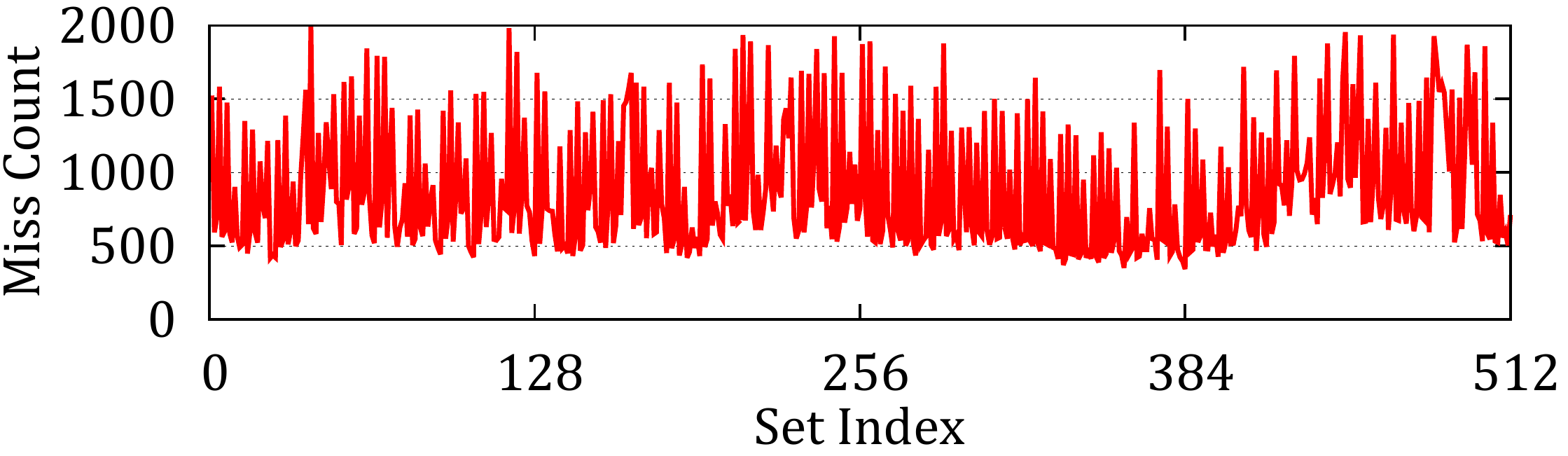}\vspace{-3pt}
        \caption{\emph{gcc}}
    \end{subfigure}
    ~ ~ ~
    \begin{subfigure}[t]{0.47\textwidth}
        \centering
        \includegraphics[width=\textwidth]{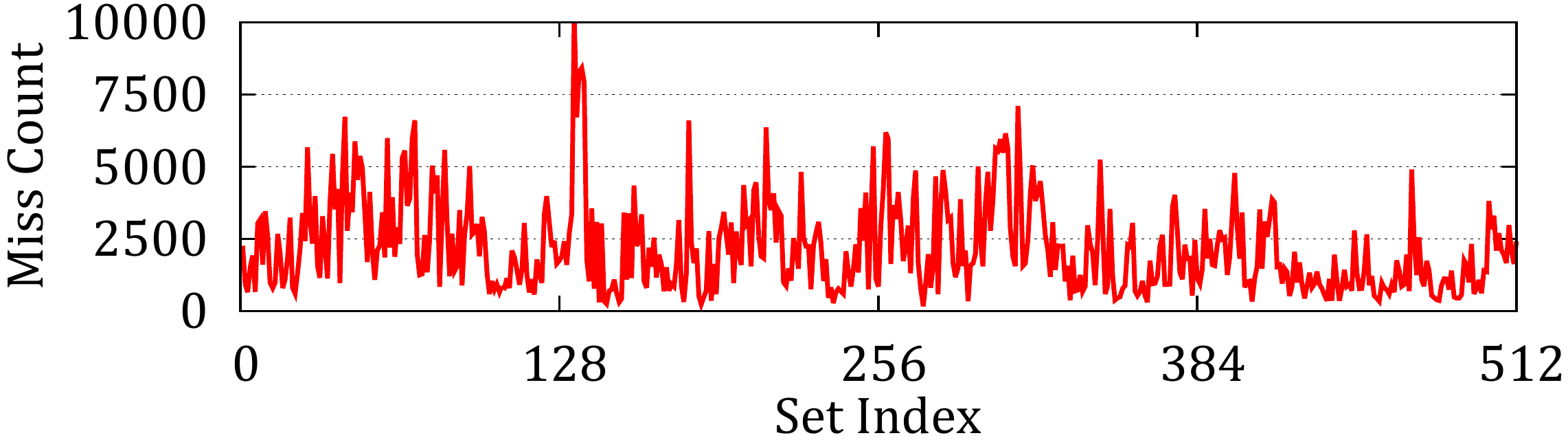}\vspace{-3pt}
        \caption{\emph{xalancbmk}}
    \end{subfigure}
    ~ ~ ~
    \begin{subfigure}[t]{0.47\textwidth}
        \centering
        \includegraphics[width=\textwidth]{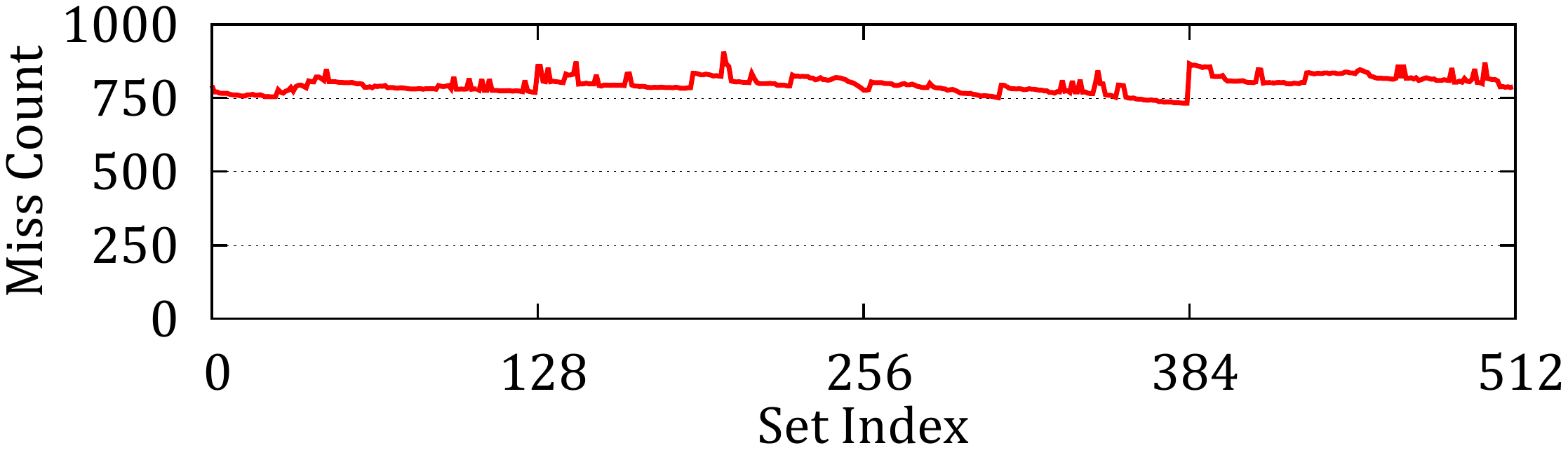}\vspace{-3pt}
        \caption{\emph{hmmer}}
    \end{subfigure}
    ~ ~ ~
    \begin{subfigure}[t]{0.47\textwidth}
        \centering
        \includegraphics[width=\textwidth]{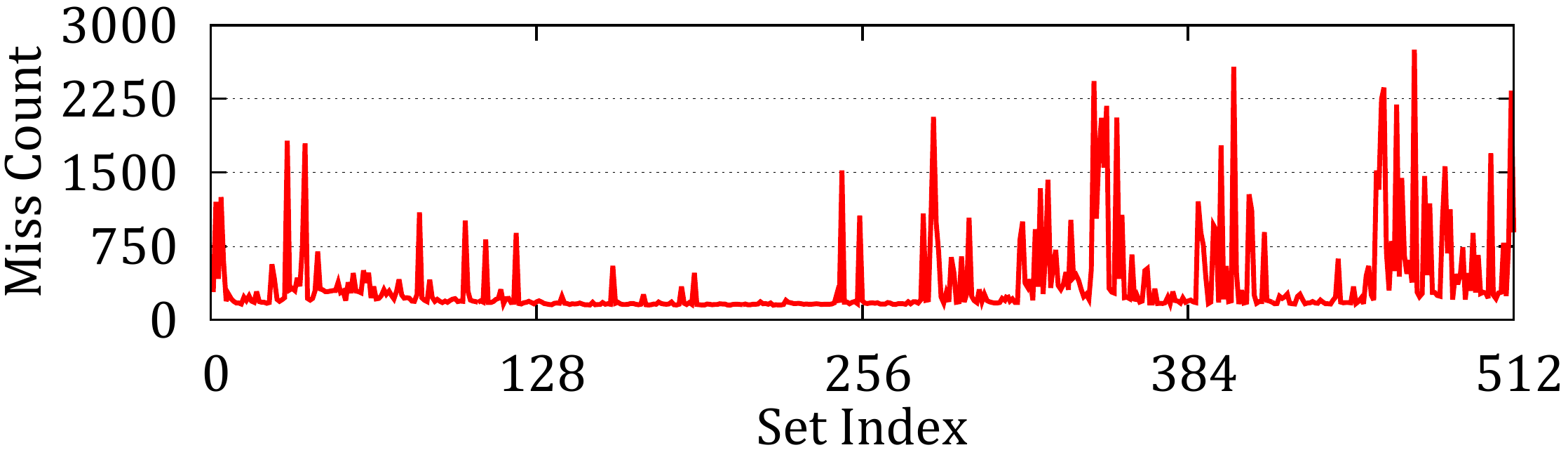}\vspace{-3pt}
        \caption{\emph{omnetpp}}
    \end{subfigure}    \vspace{-8pt}
    \caption{Distribution of the missed accesses over LLC's sets for 200 million instructions in four applications from the SPEC-CPU 2006 benchmark suite. We see that, for each application, there are some sets that have few conflict misses, while some others are very stressed.}
    \label{fig:eval2}\vspace{-13pt}
\end{figure*}

Based on this discussion, one can conclude that the MLC STT-RAM cache is not always beneficial -- it can be harmful to the cache latency and energy in certain cases. Therefore, we need a mechanism to ``dynamically'' shape-shift the MLC cache to SLC cache or vise versa (depending on the applications' dynamic cache requirements) in order to consistently achieve low memory access latency.

\subsection{Enhancements for the Stripped MLC Cache}

\subsubsection{The Need for Dynamic Associativity}
\label{Dynamic-Need}
It is not performance beneficial to shape-shift the cache configuration from SLC to MLC (or vice versa) at the granularity of an entire cache size. 
Memory references in general purpose applications are often non-uniformly distributed ac\-ross the sets of a set-associative cache.
This non-uniformity creates a heavy demand on some sets, which can lead to a high number of local conflict misses, while lines in some other sets can remain underutilized.
This fact, which is the base of our proposal, can be illustrated with some examples.
Figure~\ref{fig:eval2} plots the absolute number of conflict misses that each set exhibits in a 8-way 512KB MLC-based cache in a single-core system during one-second execution of four workloads (the same as workloads as in Figure~\ref{fig:eval2}).
For each workload, one can see that there are some sets that have few conflict misses, while some others are very much stressed.
Moreover, the numbers of these opposite-behaving sets vary from one program phase to another.
This non-uniformity of miss counts across cache sets indicates that, in the stripped cache, high associativity not only brings no benefits for sets with low-utilized lines, but it can also result in great degradation in performance, lifetime and energy consumption of the cache.
As a result, the new cache architecture we propose involves, besides the explained stripped data-to-cell mapping, an \emph{on-demand associativity policy} which dynamically modulates the associativity of each set.


To this end, we propose an \emph{on-demand associativity adjustment policy} which determines the associativity of sets in the stripped cache considering their local miss rates.
We initialize the associativity to the lowest level, which corresponds to half of the full capacity in our case.
Then, the associativity of a set will grow and decrease overtime depending on the dynamic utilization.
To mitigate the effects of slow reads and high-energy writes, when a cache line needs to be turned off, an FRHE and SRLE pair is merged into an SLC line, which uses exclusively the \emph{soft-domains}, while all \emph{hard-domains} are fixed at the same value (`0' or `1')\footnote{In our experiments, the \emph{hard-domains} are set to logic `0' since bits ``00'' and ``01'' have less overlap under severe process variation~\cite{Zhang:MLC-ICCAD}.}.
As a result, an SLC line will be read in a single cycle since the \emph{hard-domains} are known and a single resistance reference will be required.
This causes an SLC line to feature fast reads and low-energy writes.

We first describe how the decision to grow the associativity is taken.
As replacement policies are not ideal, it is clearly that the associativity should not be prematurely increased after every miss.
To achieve better cache performance, we introduce two \textit{saturation counters} for each set: a miss counter (Mcnt) and a weight counter (Wcnt).
The miss counter captures the number of misses that a set exhibits and the weight counter prevents the effects of short-term variations in misses and makes a set with large weight value to be less likely to increase its associativity.
Wcnt reflects the associativity of a set and is initialized to the minimum associativity.
In each program epoch, Mcnt is initialized to Wcnt$\times$N. 
When Mcnt reaches zero, the hardware increases the associativity by one and on the next miss, the fetched memory block will be placed into the newly-allocated way.
In an effort to balance the wear between blocks and maximize lifetime, we introduce a circular pointer (a 3-bit counter for the eight cache-line pairs of our stripped cache) indicating which cache-line pair should be selected for the next increase in associativity.
Therefore, cache lines are switched to the MLC mode in a round-robin fashion and writes are well distributed among MLC cache lines.

At the end of an epoch, Mcnt is compared with \emph{SLC-Associativity}$\times$N to decide whether the cache set associativity should be reduced.
If Mcnt is larger than \emph{SLC-Associativity} $\times$N, this indicates that the utilization is low enough to reduce the associativity by one for the next epoch.
In this case, the replacement policy is triggered and the associativity is reduced by evicting a cache line and converting the corresponding cache-line pair into SLC.

\subsubsection{The Need for a Cache Line Swapping Policy}
Besides performance efficiency, the stripped data-to-cell mapping provides an opportunity to further enhance the performance efficiency, energy efficiency, and lifetime of a MLC-based STT-RAM cache. 
More accurately, fast read lines (FRHE) greatly speed up read operations compared to the stacked mapping. On the other hand, if write-dominated lines can be directed to low write-energy lines (SRLE), the write energy and cell lifetime can be kept close to those of SLC.
To maximize the benefits provided by \emph{stripping}, we propose a swapping policy to dynamically promote write-dominated data blocks to SRLE lines and read-dominated ones to FRHE lines.
\begin{figure*}[!t]
  \center
  \includegraphics[width=0.9\textwidth]{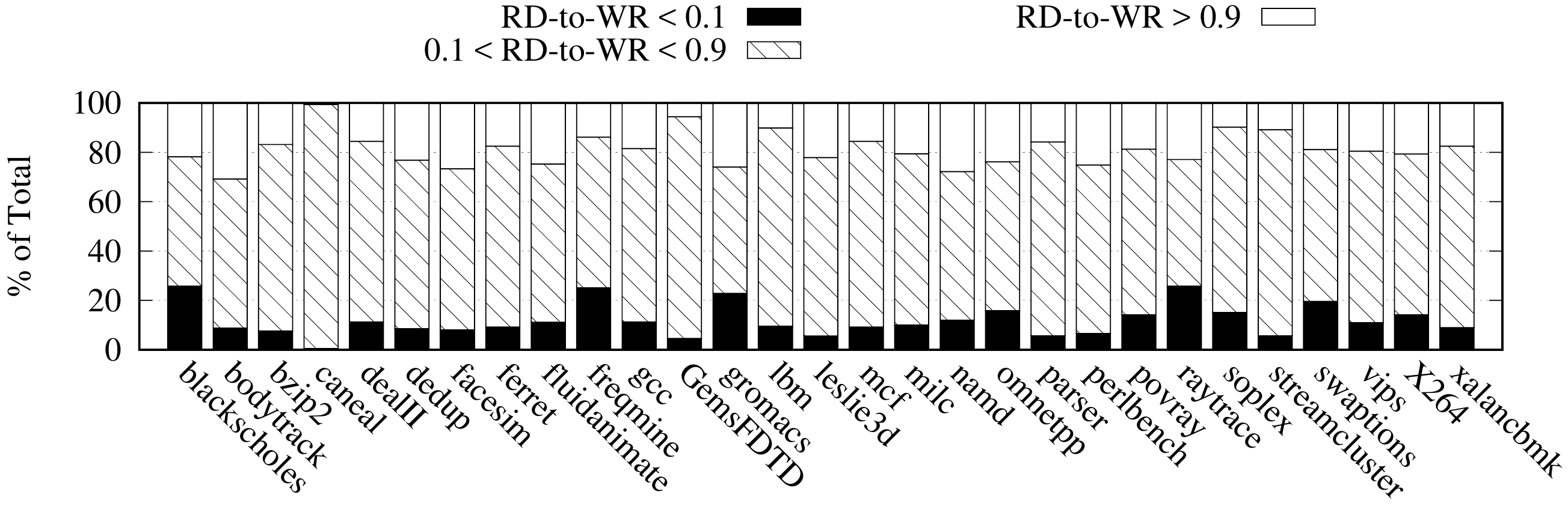}
  \caption{Percent of memory blocks in cache with read-dominated, write-dominated, and non-dominated properties for a set of workloads from PARSEC-2 and SPEC-CPU 2006 programs.}
  \label{fig-RD-WR}
\end{figure*}
The swapping mechanism in the stripped mapping must be used with care.
Looking into the cache traces, we find many blocks that are both read- and write-dominated.
In Figure~\ref{fig-RD-WR}, we report the read/write intensity of the all memory blocks in LLC for a set of workloads from the multi-thread PARSEC-2 benchmark programs~\cite{bienia09:parsec2} and the SPEC-CPU 2006 benchmark programs~\cite{Spradling07:SPECCPU06}.
If we consider that a memory block written (or read) in LLC 90\% of the time is write-dominated (or read-dominated), we observe that 32.7\% of such memory block on average, and down to 2\% for some programs.
In this situation, even utilizing a near-optimal data swap, the remaining non-dominated memory blocks might keep swapping between the FRHE and SRLE lines.
This might incur more cache contention and consume more energy.
Worse, the cache lifetime would be considerably reduced by the write amplification provoked by the swaps.
Therefore, to avoid this scenario, a swap policy is introduced.
For read- and write-aware swap, each cache line is associated with a swap counter (Scnt) and a swap weight counter (SWcnt).
At each epoch, SWcnt is initialized to `1', increments when a swap happens, and saturates when it reaches M.
Scnt is initialized at each epoch with SWcnt$\times$N, and decrements when its FRHE line is written, or when its SRLE line is read.
If Scnt reaches zero either its SRLE or FRHE line will replace the block to be evicted based on the replacement policy in use.
If the victim line is an FRHE, it is replaced by the SRLE; otherwise it is replaced by the FRHE.
Here, M is the maximum value that a hardware counter can represent (in our experiments 256 for an 8-bit counter), and N is the swap threshold that we study in the experiments.
It is to be noted that, on a miss, Scnt and SWcnt are reinitialized.

\section{Experimental Methodology}
\label{Experiments}
In this section, we describe our simulation platform as well as the design methodology used for our evaluation throughout this paper.

\subsection{Infrastructure}
\label{infrastructure}
We perform a microarchitectural, execution-driven simulation of an out-of-order processor model with Alpha ISA using the Gem5 simulator~\cite{gem5}.
The simulated CMP runs at 2.5~GHz frequency.
We use McPAT~\cite{Li09:McPAT} to obtain the timing, area, energy and thermal estimation for the CMP we model, and use CACTI~6.5~\cite{Muralimanohar07:CACTI6.5} for detailed cache area, power and timing models.
For STT-RAM LLC, NVSim~\cite{Dong:NVSim} is used and is parametrized with the cell latency and energy parameters from Table~\ref{table-model}.
We use 45~nm ITRS models, with a \emph{High-Performance} (HP) process for all the components of the chip except for LLC, which uses a \emph{Low-Operating-Power} (LOP) process.

\begin{table}[!b]
  \centering
  \scriptsize
  \caption{Main characteristics of our simulated CMPs.
The latencies shown assume a 32~nm process at 2~GHz.} \vspace{-5pt}
  \label{table-system}
  \scriptsize
  \begin{tabular}{|p{0.55in}||p{2.4in}|}
    \hline
    \multicolumn{2}{|c|}{\textbf{Processor Layer (Tier 1)}} \\
    \hline
    Cores     & 8-cores, SPARC-III ISA, out-of-order, 2~GHz, Solaris~10 OS \\
    L1 caches & 32~kB private, 4-way, 64~B, LRU, Write through, 2-port, 1-cycle Access time, MSHR: 4 instruction \& 32 data\\
    L2 cache  & 2~MB private, NUCA, Unified, Inclusive, 16-way, 64~B, LRU, Write-back, 10 cycle, MSHR: 32 (instruction and data) \\
    Coherency & MOESI directory; 2$\times$4 grid packet switched NoC; XY routing; 1-cycle router; 1-cycle link.\\
    \hline
    \multicolumn{2}{|c|}{\textbf{L3 Cache Layer (Tier 2)}} \\
    \hline
    L3 cache & NUCA, 8 banks, Shared, Inclusive, STT-RAM, 64B, LRU, Write-back, 1-port, 32$\times$64~B write buffer, 8$\times$64~B read buffer, 4-cycle L2-to-L3 latency, MSHR: 128 (instruction and data)\\
    L3 Config. & cf. Table~\ref{table-LLC-model} \\
    \hline
    \multicolumn{2}{|c|}{\textbf{Off-Chip Main Memory}} \\
    \hline
    Controller & 4 on-chip, FR-FCFS scheduling policy \\
    DRAM & DDR3 1333~MHz (MICRON), 8~B data bus, \textsuperscript{t}RP-\textsuperscript{t}RCD-CL: 15-15-15~ns, 8 DRAM banks, 16~kB row buffer per bank, Row buffer hit: 36~ns, Row buffer miss: 66~ns \\
    \hline
  \end{tabular}
\end{table}

\begin{table*}[!ht]
  \centering
  \caption{Characteristics of the evaluated workloads.}
  \scriptsize
  \label{table-Workload}
  \begin{tabular}{|lcc||lcc|}
  \hline
    \textbf{Workload} & \textbf{MPKI} & \textbf{HPKI} & \textbf{Workload} & \textbf{MPKI} & \textbf{HPKI}\\
    \hline
    \multicolumn{3}{|c||}{\emph{Multi-Threaded PARSEC-2 Workloads (8 Threads)}} & \multicolumn{3}{c|}{\emph{8-Application Multi-Programmed (SPEC-CPU 2006)}}\\ \hline
    blackscholes & {0.78 (L)} & {37.33 (H)} & MP1: 2 Copies of (xalancbmk,omnetpp,bzip2,mcf) & {20.16 (H)} & {39.51 (H)} \\
    bodytrack & {0.96 (L)} & {11.90 (M)} & MP2: 2 Copies of (milc,leslie3d,GemsFDTD,lbm)  & {33.01 (H)} & {38.69 (H)} \\
    canneal & {15.19 (H)} & {27.13 (H)} & MP3: 2 Copies of (mcf,xalancbmk,GemsFDTD,lbm)  & {24.23 (H)} & {37.55 (H)} \\
    dedup & {3.04 (M)} & {9.072 (M)} & MP4: 2 Copies of (mcf,GemsFDTD,povray,perlbench)  & {14.89 (H)} & {23.41 (H)} \\
    facesim & {10.66 (H)} & {14.26 (M)} & MP5: 2 Copies of (mcf,xalancbmk,perlbench,gcc) & {18.33 (H)} & {49.41 (H)} \\
    ferret & {7.80 (M)} & {23.21 (M)} & MP6: 2 Copies of (GemsFDTD,lbm,povray,namd) & {6.99 (M)} & {11.68 (M)} \\
    fluidanimate & {5.54 (M)} & {10.51 (M)} & MP7: 2 Copies of (gromacs,namd,dealII,povray) & {1.85 (M)} & {7.94 (M)} \\
freqmine & {0.51 (L)} & {7.30 (M)} & MP8: 2 Copies of (perlbench,gcc,dealII,povray) & {5.21 (M)} & {25.63 (H)} \\
raytrace & {0.45 (L)} & {0.92 (L)} & MP9: 2 Copies of (namd,povray,perlbench,gcc) & {2.22 (M)} & {8.87 (M)} \\
streamcluster & {0.51 (L)} & {5.35 (M)} & MP10: 2 Copies of (milc,soplex,bzip2,mcf) & {22.94 (H)} & {38.89 (H)} \\
swaptions & {0.15 (L)} & {4.34 (M)} & MP11: 2 Copies of (parser,gcc,namd,povray) & {4.27 (M)} & {26.39 (H)} \\
vips & {2.24 (M)} & {15.69 (M)} &  & & \\
x264 & {1.22 (M)} & {12.92 (M)} &  & & \\
\hline
  \end{tabular}
\end{table*}

\begin{table}[!t]
  \centering
  \scriptsize
  \caption{Evaluated LLC configurations.}
  \label{table-LLC-model}
  \begin{tabular}{|c||r@{\ }l|r@{\ }l|c|}
  \hline
    \multirow{2}{*}{\textbf{L3 Config.}} & \multicolumn{2}{c|}{\textbf{Latency}} & \multicolumn{2}{c|}{\textbf{Dynamic}} & \textbf{Leakage} \\
    & \multicolumn{2}{c|}{\textbf{[cycles]}} & \multicolumn{2}{c|}{\textbf{Energy [nJ]}} & \textbf{Power [W]} \\
    \hline
    8~MB   & Lookup: & 3       & hard-R: &0.34 & {1.52} \\
    8-to-16way    & hard-R-hit: &3   & soft-R: &0.38 & \\
     stripped MLC                 & soft-R-hit: &5   & hard-W: &1.93 & \\
                     & hard-W-hit: &19  & soft-W: &1.28 & \\
                     & soft-W-hit: &42  &              & & \\
    \hline
    5MB 8way     & Lookup: &1       & R: &0.32      & {0.156} \\
      SLC                & R-hit: &3        & W: &1.29      & \\
                     & W-hit: &19       & &             & \\
    \hline
    8MB 16way        & Lookup:& 3       & R: &0.64      & {0.152} \\
    stacked MLC      & R-hit: &5        & W: &1.58      & \\
                     & W-hit: &37       &  &            & \\
    \hline
    8MB 16way     & Lookup:& 2       & R: &0.32      & {0.217} \\
     SLC                & R-hit: &3        & W: &1.29      & \\
                     & W-hit: &19       & &             & \\
    \hline
  \end{tabular}
\end{table}

\subsection{System}
\label{System}
We model the 8-core CMP system detailed in Table~\ref{table-system}.
The system has three levels of caches and, because STT-RAM is not compatible with CMOS, it is built on a 2-tier 3D integration.
At the processor tier (near to the heat sink), the cache hierarchy has a split L1 private instruction and data cache for each core.
Each core also has a private L2 cache that is kept exclusive to the L1 cache.
The STT-RAM L3 cache (LLC) is \textit{logically shared} among all the cores while physically structured as static NUCA and mounted at the top tier of 3D die~\cite{Sun09:STT-RAM,Black06:3D,Mishra11:3D-STT-RAM}.
On 45~nm, the CMP area is estimated to 60~$mm^2$ and has a TDP of 77~W at 2~GHz and 1.1~V supply voltage.

Using McPAT~\cite{Li09:McPAT}, the processor layer in our 3D IC (i.e., tier~1) has a 5.1~mm\textsuperscript{2} die area.
By assuming an SLC STT-RAM cell size of 14~F\textsuperscript{2}~\cite{Sun09:STT-RAM}, we derived that a 5~MB SLC can fit in same die area in tier~2.
Since the area of an STT-RAM cell is dominated by its access transistor and we use an SLC tag array for our MLC data array, 8~MB MLC STT-RAM can fit within the 5.1~mm\textsuperscript{2} area.
Table~\ref{table-LLC-model} summarizes the configuration of the LLC in the proposed system as well as in three reference configurations: a 5~MB SLC cache (\emph{fast cache}), an 8~MB 2-bit MLC cache with stacked mapping(\emph{dense cache}), both with the same die area, and an 8~MB SLC cache (fast-dense cache) with double area.

The performance of static STT-RAM NUCA, is highly sensitive to write operations that can be blocking for the subsequent read requests.
To alleviate this inefficiency, each bank has a separate 8-entry \emph{Read Queue} (RDQ) and 32-entry \emph{Write Queue} (WRQ) that queue all pending requests.
A read request to a line pending in the WRQ is serviced by the WRQ.
When a bank is idle and either RDQ or WRQ (but not both) is non-empty, the oldest request from that queue is serviced.
If both RDQ and WRQ are non-empty, then a read request is serviced unless the WRQ is more than 80\% full, in which case a write request is serviced.
This ensures that read requests are given priority for service in the common case, and write requests eventually have a chance to get serviced.

\subsection{Workloads} 
For multi-threaded workloads, we use the complete set of parallel programs in PARSEC-2 suite~\cite{bienia09:parsec2}.
For multi-program evaluation, we use the SPECCPU2006 benchmarks.
We classify a benchmark as \emph{memory-intensive} if its L3 cache \emph{Misses Per 1~K Instruction} (MPKI) is greater than three; otherwise, we refer to it as \emph{memory non-intensive}.
Also, we say a benchmark has \emph{cache locality} if the number of L3 cache \emph{Hits Per 1K Instruction} (HPKI) for the benchmark is greater than five.
Each benchmark is classified by measuring the hits and misses when running alone in the 8-core system given in Table~\ref{table-system}.
For the multi-program workload selection, we used eleven 8-core-application workloads that are chosen such that each workload consists of at least six \emph{memory-intensive} applications and two applications with good \emph{cache locality}.

\begin{figure*}[!t]
  \center
  \includegraphics[width=0.9\textwidth]{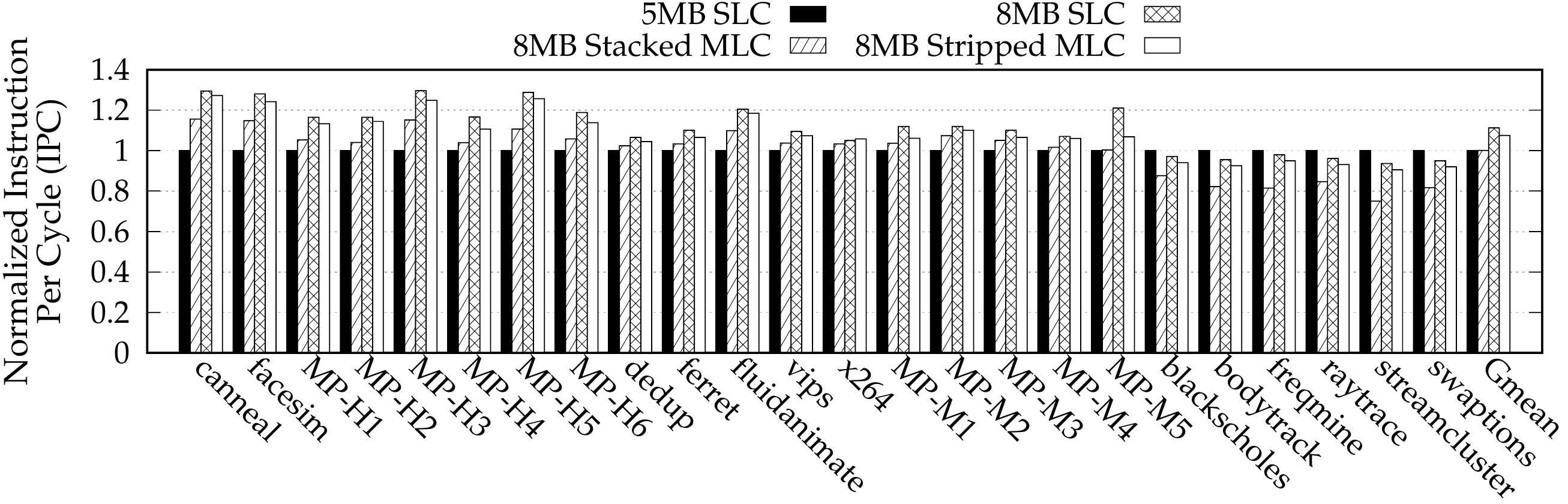}
  \vspace{-10pt}
  \caption{Percentage of IPC improvement for the proposed cache architecture with respect to the baselines. The
  proposed architecture has capacity advantage of MLCs in applications with high misses (9 first) programs and medium misses (13 second) programs.
  It also has the SLC access latency in 8 applications with low misses (at left side).}
\vspace{-5pt}
  \label{fig-IPC}
\end{figure*}

\begin{figure*}[!t]
  \center
  \includegraphics[width=0.9\textwidth]{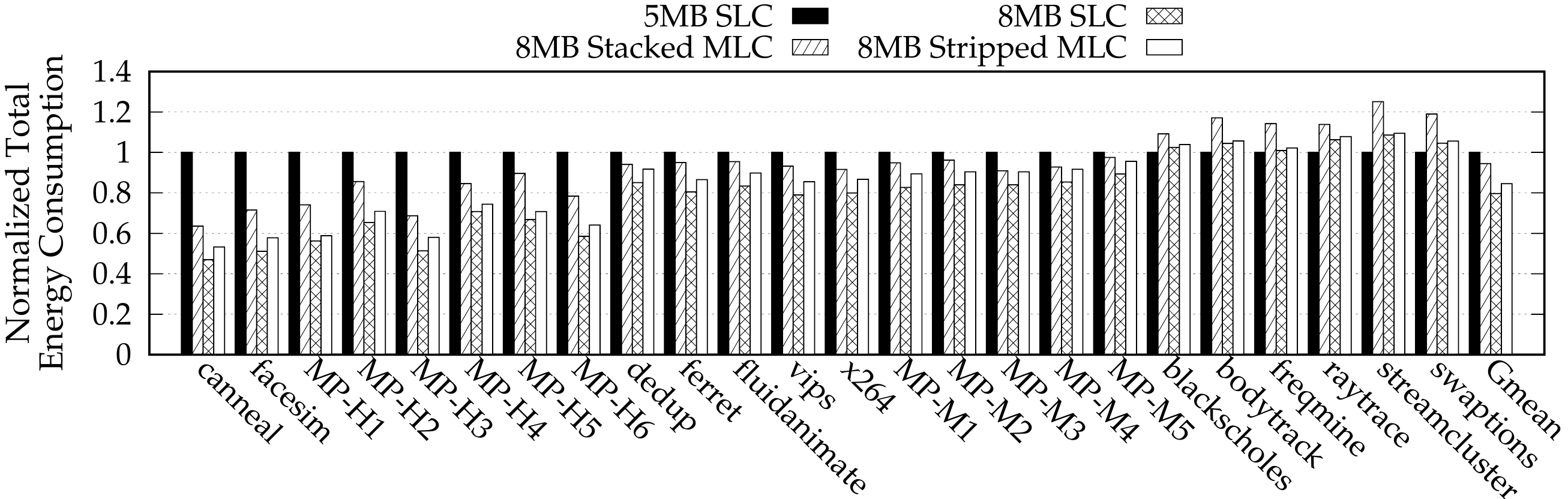}
  \vspace{-10pt}
  \caption{Total energy consumption of the cache architectures normalized to the SLC baselines. It shows that
  proposed architecture uses low read and low write energy of FRHE and SRLE lines.}
  \label{fig-energy}
\end{figure*}

Each application is simulated to completion and the results are taken from the instructions in parallel region (i.e., \emph{the region of interest}).
Regarding the input sets, we use Large set for the PARSEC-2 applications and sim-large for the SPECCPU2006 workloads.
All applications are compiled using ICC (Intel C Compiler) and IFORT (Intel Fortran Compiler) at the O3 optimization level.
Table~\ref{table-Workload} characterizes the evaluated workloads based on L3 MPKI and HPKI for the SLC reference system of Table~\ref{table-system}.
To justify the evaluation results, our workloads are classified based on their L3 miss count and hit count intensity: considering L3 MPKI, each workload is either high-missed (H) if MPKI is greater than 10, medium-missed (M) if MPKI is between 1 and 10 or low-missed (L) if MPKI is less than 1.
A workload is either high-hit (H) if HPKI is greater than 20, medium-hit (M) if HPKI is between 1 and 20 or low-hit (L) if HPKI is less than 1.

\section{Evaluation Results}
\label{Results}
Our dynamic stripped cache has a minimum and maximum associativity of 8 ways and 16 ways, respectively.
The first and second baselines are the SLC and stacked MLC cache with the same die area and line size, but different associativities: 8 ways for SLC baseline and 16 ways for stacked MLC.
The last baseline is a cache with SLC devices with the same capacity and associativity of MLC, but the die area is doubled.
For our dynamic configuration, on the other hand, a read hit or a write hit can be serviced by either MSB lines (i.e., MSB read bit or LSB read hit) or LSB lines (i.e., MSB write hit or LSB write miss).

The proposed cache organization centers around the use of SLC devices in applications with low misses and MLC devices in applications with medium and high misses.
Ultimately, this mechanism attempts to reduce the miss penalty by the same measures as an MLC cache.
Thus, an upper bound on the miss reduction for the proposed mechanism is provided by stacked MLC cache of the same size.
Our cache is expected to approach this upper bound for high missed and medium missed applications.
This upper bound can result in our scheme outperforming the SLC baseline as seen in the results.
The two other upper bound for the proposed cache are determined by the read hit at fast read lines (i.e.,
FRHE) and a write hit at low write-energy lines (i.e.,
SRLE).
These second and third upper bounds determine the reduction in access latency compared to MLC arrays and is provided by the SLC baseline with double capacity (last baseline).

\subsection{Performance Evaluation}
For programs with the high and medium L3 MPKI, we expect a higher effect on the latency when
increasing cache associativity. This is also observed in Figure \ref{fig-IPC} that plots the CPI improvement for a
system with proposed cache structure support with respect to the studied baselines. For each benchmark, the results are
\textit{normalized} to the SLC baseline for ease of comparison. This figure shows an improvement of up to 29\% in CPI of the high
associativity caches (i.e., MLC cache configurations and 8~MB SLC cache) with respect to the 4~MB SLC baseline. Our scheme also
outperforms the 8~MB stacked 2-bit cache by 10\% on average thanks to being able to construct FRHE and SRLE lines without
generally loosing maximum way associativity requirement of a set. Comparing the results with the 8~MB SLC cache baseline,
it can be seen that the performance of the system with proposed cache structure is within 5\% of the maximum performance observed. 

For applications
with low miss ratio in the LLC, one can see our cache configuration behave like an SLC baseline in most applications. Only in
some application, there is a slight degradation in overall system performance (up to 4\% in \emph{blackscholes}) that is
because of the higher latency of 8~MB NUCA access circuit compared to 4~MB SLC configuration. In short, as the last group of bars
show in Figure~\ref{fig-IPC}, the cache architecture with dynamic associativity achieves 10\% CPI improvement (on average) for
all applications with different miss ratio behavior.

\begin{figure*}[!t]
  \center
  \includegraphics[width=0.9\textwidth]{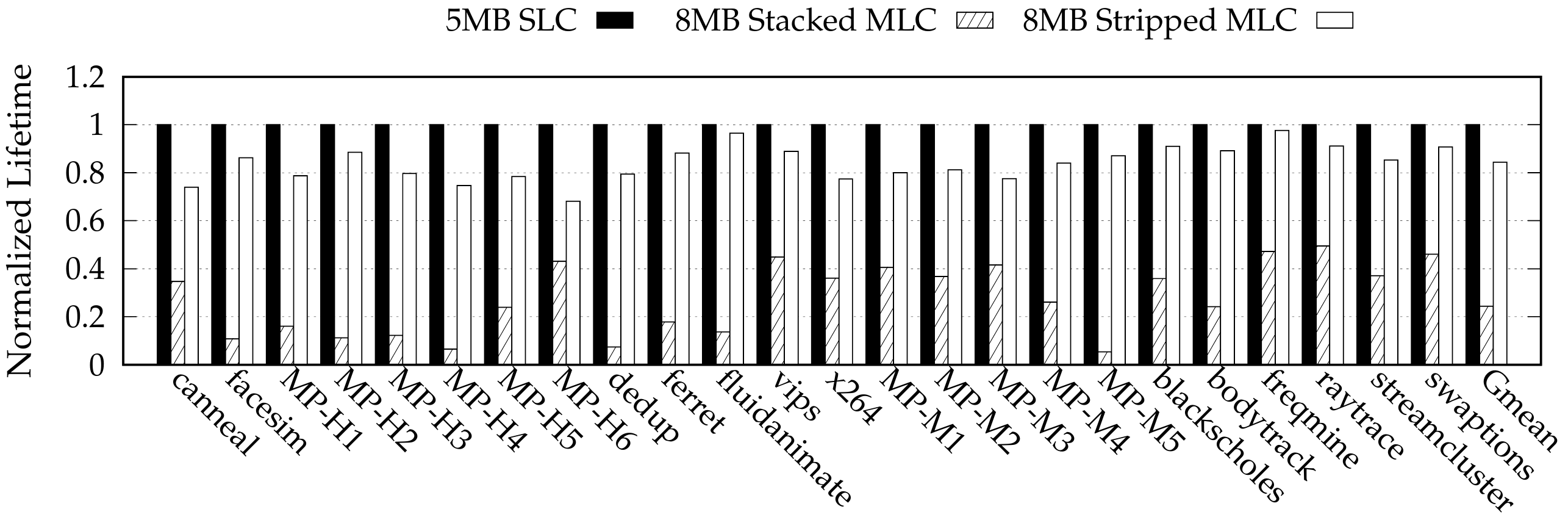}
  \vspace{-10pt}
  \caption{Lifetime of the cache configurations normalized to the SLC baselines. Our, proposed stripped scheme tries to act like
  SLCs to reach maximum lifetime.}
  \label{fig-lifetime}
\end{figure*}

\subsection{Energy Consumption}
The percentage of reduction in total memory energy compared to the baseline STT-RAM cache configurations is shown in Figure~\ref{fig-energy}. This evaluation includes the energy consumptions of both the STT-RAM LLC and off-chip main memory, and uses the
energy model given in Table~\ref{table-LLC-model}. Generally speaking, compared to the SLC baselines and the MLC baseline, the
percentage reduction in energy consumption follows the same trend observed for the system performance improvement. In other words,
we can see that the energy consumption of the proposed scheme is much better than the SLC baseline in all applications with high and
medium misses due to the higher hit ratio of the on-chip memory hierarchy. On the other hand, the energy consumption of our scheme is
better than the MLC baseline with the stacked data-to-cell mapping, as it constructs lines with low write energy and trying to
allocate them write-dominated blocks. Also, with respect to the SLC baseline cache, the proposed cache
architecture results in 17\% reduction in the total memory energy on average (and 29\% to 46\% for programs such as \emph{canneal},
\emph{facesim}, and \emph{art} with high MPKI).

\subsection{Lifetime Evaluation}
In this work, we assume that reliable writes into an SLC STT-RAM cell is limited to 10\textsuperscript{12} cycles
\cite{Sun09:STT-RAM}, and it is linearly scaled down for 2-bit STT-RAMs (i.e., exactly one-tenth). For the lifetime evaluation, the
main memory traces are extracted from the full-system simulator and are fed into a simulation tool. To avoid cache failure on
wear-out of limited cells, each cache line is augmented with an ECC correcting up to 5 faulty bits. After each
cache write, the read access circuit is used to read it out and a cache line (or equally a physical way) is assumed to be dead
if the read-out data block has more than 5 bit mismatches with the original data. Finally, the simulator keeps track of the write counts on
each coding set until a cache set has more than four dead physical ways. We measure this duration and use it to estimate
and analyze the lifetime. The lifetime of the complete cache architecture is compared with the 5~MB
SLC caches and an 8~MB stacked MLC cache in Figure \ref{fig-lifetime}. Overall, the proposed scheme provides a lifetime larger than 70\% of an SLC cache with identical ECC strength.

\section{Related Work}
\label{Back1}
In this section, we go over the most relevant pieces of work on STT-RAM cache memories as well as cache associativity.

\subsection{STT-RAM Cache Memories}
Owing to real concerns of SRAM power in nanometer regime, resistive memories such as STT-RAM are presented to offer a highly-scalable low-leakage alternative for large cache arrays.
Compared to competitive non-volatile memories (such as ReRAM, PcRAM and FeRAM), STT-RAM benefits from the best attributes of fast nanosecond access time, CMOS process compatibility, high density, and better write endurance.
Dong et al. give a detailed circuit-level comparison between SRAM cache and STT-RAM cache in a single-core microprocessor~\cite{Dong08:STT-RAM-Circuit}.

Based on the findings given in this study, Sun et al. extended the application of STT-RAM to NUCA cache substrate in CMPs and studied the impact of the costly write operation in STT-RAM on power and performance~\cite{SUN07:SPARC-TECHREP}.
To address the slow write speed and high write energy of STT-RAM, many proposals have been made in recent years.
Zhou et al. proposed an early write termination scheme that uses write current as a read bias current and cancels an ongoing write if the write data is unnecessary~\cite{Zhou09:EWT}.
Alternative approaches are SRAM/STT-RAM hybrid cache hierarchies and some enhancements, such as write buffering~\cite{Sun09:STT-RAM}, data migration~\cite{Sun09:STT-RAM,Arjomand11:STT-RAM,Wu09:STT-RAM}, and data encoding~\cite{Arjomand12:STT-RAM}.
As a cache solution with uniform technology, previous work proposed to trade off the non-volatility of STT-RAM for write performance and power improvement~\cite{Smullen11:Retention,Sun11:STT-RAM-Retention,Jog12:Cache-Revive}.
To ensure data integrity in these architectures, some DRAM-style refresh schemes are introduced which may not scale well for large cache capacities. 

Regarding MLC STT-RAM, Chen et al. proposed a dense cache architecture using devices with parallel MTJs~\cite{Chen12:Hybrid-STT-RAM}.
Although they use the MLCs with parallel MTJs to have lower write power (compared to series MTJs) suitable for cache~\cite{Chen10:MLC-STT}, a reliability comparison of these two devices show that parallel devices confront serious challenges in nanometer technologies with large process variations~\cite{Zhang:MLC-ICCAD}.
Finally, some recent proposals studied the effect of decoupling bits of an MLC device (STT-RAM~\cite{Jiang12:MLC-DAC} or PcRAM~\cite{Yoon12:Decoupled-MLC}) in performance, energy, and reliability improvement of non-volatile memories.

\subsection{Reducing Conflict Misses in Caches}
There are several prior works targeting conflict misses in set-associative caches, which can be generally categorized as follows:

\textbf{Category 1: Hashing block address --} Instead of using a subset of the address bits to index the cache, one can use a better hash function on the address to reduce conflict misses by spreading out accesses. Although hashing relaxes the miss pressure on some sets with large working set~\cite{Kharbutli04:PIC}, it slightly increases access latency as well as area and power overheads due to this additional circuitry. Moreover, it needs to store full block address in tag which adds to tag store overheads.

\textbf{Category 2: Skew-associative caches --} Skew-associ\-ative cache~\cite{Seznec93:Skewed} indexes each way with a different hash function. Then, a block address conflicts with a fixed set of blocks, but those blocks conflict with other addresses. This spreads out the conflicts.
Based on skew-associative design, ZCache~\cite{Sanchez10:Zcache} decouples the notion of ways and associativity.
Ways represent the number of tags that must be searched when looking up a cache line, while associativity is referred as the number of blocks that can be evicted to make room for an incoming line. The ZCache keeps the number of ways small, but it has a large associativity.
Although skew-associative caches typically exhibit lower conflict misses than a set-associative cache with the same number of ways~\cite{Bodin95:Skew-Performance}, they break the concept of a set, so they cannot use replacement policy implementations that rely on set ordering.

\textbf{Category 3: Mapping multiple locations to one physical way -- } Column-associative caches~\cite{Agarwal93:CSAC} extend the concept of direct-mapped caches
to allow a block to reside in two locations based on a primary and a secondary hash functions. In this approach, tag lookup is a two-step mechanism. First, it uses the primary hash function to search for a block, and second, it uses the secondary function on the cache miss.
To improve access latency, a hit on the second step triggers a swapping logic to reorder first and second search functions. Similar techniques predict which location to be searched first~\cite{Calder96:PSA}, and there are also the schemes that try merging the less used sets with more used ones to balance utilization of the sets~\cite{Rolan09:SBC}.

The main shortcomings of these approaches are the variable hit latency, reduced cache bandwidth due to multiple lookups, and additional energy required to do swaps on hit accesses.

\textbf{Category 4: Victim cache --} A victim cache is a fully-associative small cache that keeps blocks evicted from the main cache until for future reuses~\cite{Jouppi90:Victim}.
It can avoid conflict misses that are re-referenced after a short period. However, due to its limited capacity, it is not particularly useful when the number of sets with large local misses are considerably large~\cite{Brehop03:Math-for-cache}.
Inspired by this scheme, Scavenger~\cite{Basu07:Scavenger} divides cache space into two equally large parts, a conventional set-associative cache and a victim cache organized as a heap.
Although victim cache results in better performance in applications with moderate miss ratios, it suffers from the additional latency and energy consumption needed for checking the victim cache, regardless of the hit or miss on the victim cache.

\textbf{Category 5: Employing pointer-like tag array -- } An alternative strategy is to implement tag and data arrays separately, making the tag array highly associative, and using it as pointers to an array of data blocks. Examples in this line are Indirect Index Cache (IIC)~\cite{Hallnor00:FAS} and V-Way cache~\cite{Qureshi05:V-Way}. IIC implements the tag array as a hash table using open-chained hashing for high associativity. The V-Way cache, on the other hand, implements a conventional set-associative tag array, but makes it larger than the tag array to reduce conflict misses.
Tag indirection schemes suffer from two problems. First, they usually increase hit latency, as they have to serialize tag lookup and data access. Second, the tag array overheads in these two schemes are large (around 2$\times$), and may not be acceptable for large cache arrays.

\begin{figure}[!t]
  \center
  \includegraphics[width=\columnwidth]{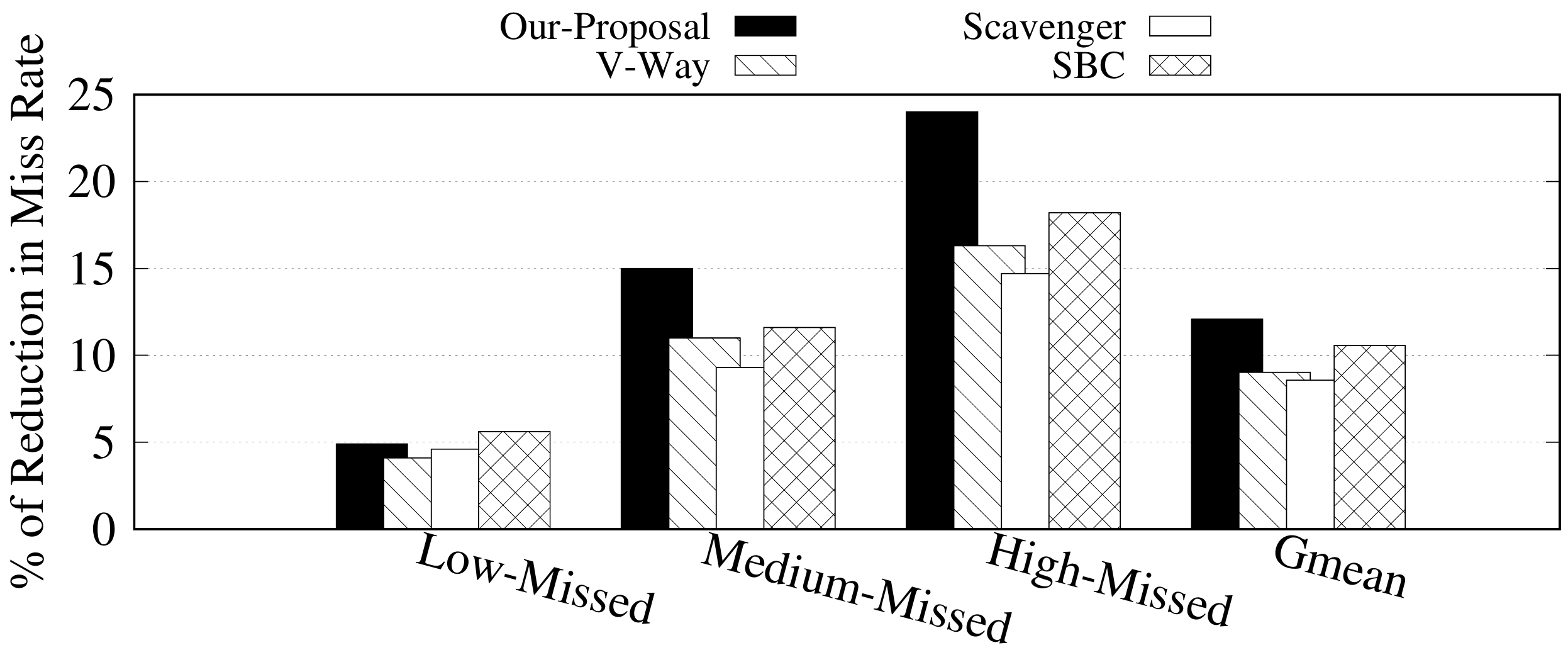}
  \vspace{-10pt}
  \caption{Comparison of our proposed cache with V-Way~\cite{Qureshi05:V-Way}, Scavenger~\cite{Basu07:Scavenger}, and SBC~\cite{Rolan09:SBC} caches in terms of the percentage reduction in cache misses relative to the SLC cache in previous configuration.
Note that cache sizes are set to have same die area.
This figure shows that our technique is better than its counterparts in miss ratio, especially when the application requires large associativity.}
  \label{fig-prior-tech}
\end{figure}

\subsubsection{Comparative Analysis}
Several proposals increase cache associativity relying on techniques requiring either heaps~\cite{Basu07:Scavenger}, hash table~\cite{Hallnor00:FAS} or prediction mechanism~\cite{Calder96:PSA}.
This may increase energy and latency of cache hits and the resulting cache design may be much more complex than conventional cache arrays.
On the other hand, this paper focuses on multi-bit capability of cutting-edge STT-RAM technology and proposes a workload-aware per-set associativity regulation using its SLC to MLC (and viceversa) shapeshifting property.

Here, we compare the proposed MLC STT-RAM cache system against state-of-the-art works on cache associativity applied to the same platform (i.e., STT-RAM cache).
We use three schemes for comparison, including the \emph{V-Way cache}~\cite{Qureshi05:V-Way}, the \emph{Scavenger cache}~\cite{Basu07:Scavenger}, and the \emph{dynamic SBC cache}~\cite{Rolan09:SBC}.
For fair analysis, these three approaches are applied to SLC with the same die size.
Figure~\ref{fig-prior-tech} compares the LLC miss rates of these schemes for the configuration used previously (Table~\ref{table-system}).
The results are shown as an \textit{average reduction} in miss rate for different workloads categories in Table~\ref{table-Workload}, i.e., \emph{workloads with low LLC miss rate}, \emph{workloads with moderate miss rate}, and  \emph{workloads with high miss rate}. The results are normalized to the miss rate of the baseline SLC configuration.
We can observe that, the results vary between the 12.1\% reduction for our proposed solution and the 8.6\% reduction for Scavenger. And, SBC achieves a 10.8\% reduction, better than the 9\% obtained by V-Way.
More accurately, the proposed cache is the best one in high and medium missed category and SBC is the best for workloads with low miss accesses.
We must take into account that V-Way cache turns misses into hits, while the other three one turns them into secondary hits, which suffer the delay of a second access to the tag array.
On the other hand, the duplication of the tag-store entries, the addition of one pointer to each entry and a mux to choose the correct pointer increases the V-Way tag access time by around 39\%, while our solution along with SBC involves very light structures, thus having a negligible impact on access time.

\section{Conclusion}
\label{Conclusion}
The emerging technology of SLC STT-RAM has been shown to be a promising candidate for building large last-level caches. The natural next step would be to use MLC STT-SRAM, but its advantage in doubling the storage density comes with a number of serious shortcomings in terms of lifetime, performance, and energy consumption. In this paper, we have shown that, by operating MLC STT-RAM in SLC mode when the additional density is not required, one can achieve the best of both worlds and improve performance and energy with only a minimal impact on lifetime. This improvement requires more than a naive shut down of unused ways in the cache (which are thus used in SLC mode instead of MLC mode) and we have shown how one should actively migrate data across physical ways to maximize the benefits of this technique. This work shows that emerging memory technologies can be efficiently accommodated in traditional memory technologies, but they require some new techniques for the integration to be successful.

\bibliographystyle{abbrv}
\bibliography{ref}


\end{document}